\documentclass[preprint]{elsarticle}

\makeatletter
\def\ps@pprintTitle{%
 \let\@oddhead\@empty
 \let\@evenhead\@empty
 \def\@oddfoot{\centerline{\thepage}}%
 \let\@evenfoot\@oddfoot}
\makeatother

\usepackage[a4paper, left=35mm, right=35mm, top=35mm, bottom=35mm]{geometry}
\usepackage{lineno,hyperref}
\modulolinenumbers[5]

\usepackage{amsfonts}
\usepackage{amsmath}
\usepackage{amssymb}
\usepackage{xcolor}
\usepackage{multicol}
\usepackage{multirow}
\usepackage[normalem]{ ulem }
\usepackage{soul}

\begin{document}

\begin{frontmatter}

\title{A data-driven convergence criterion for iterative unfolding of smeared spectra}


\author[LPSC]{M. Licciardi\corref{mycorrespondingauthor}}

\author[LLR]{B. Quilain}
\cortext[mycorrespondingauthor]{Corresponding author, \href{mailto:licciardi@lpsc.in2p3.fr}{\tt licciardi@lpsc.in2p3.fr}}

\address[LPSC]{Univ.~Grenoble Alpes, CNRS, Grenoble INP, LPSC-IN2P3, 38000 Grenoble, France}
\address[LLR]{LLR, Ecole Polytechnique, CNRS/IN2P3, Université Paris‐Saclay, Palaiseau, France}

\begin{abstract}
A data-driven convergence criterion for the D'Agostini (Richardson-Lucy) iterative unfolding is presented. It relies on the unregularized spectrum (infinite number of iterations), and allows a safe estimation of the bias and undercoverage induced by truncating the algorithm. In addition, situations where the response matrix is not perfectly known are also discussed, and show that in most cases the unregularized spectrum is not an unbiased estimator of the true distribution.
Whenever a bias is introduced, either by truncation of by poor knowledge of the response, a way to retrieve appropriate coverage properties is proposed.
\end{abstract}


\end{frontmatter}

\section*{Introduction}
Unfolding procedures are at the heart of many domains in science and engineering, from optics to high-energy physics. These procedures aim at answering the apparently simple question: what is the true physical distribution that led to the observed data? The answer is however often not simple, since detector effects (such as finite resolution or limited acceptance) smear the signal and information about the initial distribution is partially lost. Moreover, there may be some non-trivial transformation between the true variable and the observed one: for example, one could like to infer the momentum of a particle from its penetration length in a calorimeter. Many unfolding (or deconvolution) techniques have been proposed in the last decades \cite{1972Richardson,Lucy:1974yx,Multhei:1986ps,H_cker_1996, DAgostini1995,blobel2002unfolding} to solve this statistical problem, in a large variety of physics fields. 

A general issue when solving such inverse problems is that unfolding procedures enhance fluctuations. Indeed, to counter the smearing by detector effects, unfolding techniques act as \textit{anti-smearing} processes. Any true distribution will be smeared when folded through the detector finite resolution, so we may find a true spectrum with large fluctuations while the corresponding observed data remains relatively smooth. To mitigate this effect one uses \textit{regularization} techniques to encode some additional information about, for instance, the smoothness, curvature or generic shape of the true distribution. Doing so, the variance of the true distribution's estimator is reduced, but biases are introduced, and the regularized variance do not provide proper frequentist coverage anymore. The strength of the regularization may be tuned to balance decreasing variance with increasing bias. This tuning suffers however from arbitrariness since there seem not to be many consensual prescriptions (if any), though the statistics field has proposed several criteria \cite{Lcurve,GCV}. Popular regularized methods in high-energy physics are such as penalized log-likelihood minimization or Tikhonov regularization \cite{Tikhonov:1963,H_cker_1996,blobel2002unfolding}, filtered Singular Value Decomposition \cite{TSVD,WienerSVD}, and truncated iterative unfoldings \cite{Zech_2013,zech2016analysis}. 

The iterative unfolding algorithm introduced by D'Agostini \cite{DAgostini1995,DAgostini2010} in high energy physics, and known as Richardson-Lucy \cite{1972Richardson,Lucy:1974yx} in astrophysics since the 1970s, is widely used. D'Agostini's formulation is based on Bayes' inversion formula for conditional probabilities and is interpreted in terms of bayesian statistics. This algorithm however appears to be equivalent to the expectation-minimization (EM) algorithm applied to obtain a maximum-likelihood estimator (MLE) for Poisson likelihoods (see for example a derivation in \cite{KuuselaMSc}), and can be efficiently used apart from its bayesian interpretation. 

In such EM algorithms, a initial guess of the true distribution is required at the start, but the limit point of the algorithm is an unbiased MLE. The regularization is introduced by stopping the algorithm after a small number of iterations. Doing so, the estimator is biased and depends on the initial guess that has been used to start the iterations. In analyses using this iterative algorithm, the cut-off is chosen, at best, after more or less detailed Monte-Carlo (MC) studies ensuring that the bias introduced by the truncation is reasonnably small -- in the worst case, no dedicated studies are done at all. In any situation, such a MC-based approach is only valid if MC distributions used for these tests are in close compatibility with the observed data. However, if the truncation bias is evaluated based on MC distributions that do not reproduce well the data, there is no guarantee for the chosen cut-off to be appropriate for the data set of interest, and the level of bias may be well underestimated.

In order to remedy this effect, we present here a data-driven criterion to choose an appropriate number of iterations for D'Agostini-like iterative algorithms. It provides an upper bound of the regularization bias that can be used to correct for under-coverage of the error estimates. This criterion has been primarily developped and used for neutrino-nucleus cross section measurements \cite{licci2018}. Because of nuclear collective effects, neutrino-nucleus interactions Monte-Carlo generators \cite{NEUT,GENIE} have sizeable uncertainties and their respective predictions are not always consistent with each other (though remarkable improvements has been done in the last decade). In this field, as in others where Monte-Carlo simulations are suspected to not be as accurate as expected, such a data-driven criterion will hopefully help for more cautious data analysis.

This article is organized as follows. The iterative unfolding algorithm and its properties are recalled in section~\ref{scn:presentation}. In section~\ref{scn:2peak-model}, a two-peak toy model used for illustration is introduced. Section~\ref{scn:DD-criterion} presents the convergence criterion, and coverage properties of unfolded spectra are studied. In section \ref{scn:wrong-response} the cases where the response matrix is not perfectly known are discussed. Finally, we conclude and discuss some other methods in section~\ref{scn:ccl}.

\section{Iterative unfolding}\label{scn:presentation}

The iterative unfolding algorithm aims at recovering the distribution of a true variable $X^\mathrm{true}$ provided the observation of its observed counterpart $Y^\mathrm{obs}$. We follow here the description of D'Agostini \cite{DAgostini1995}, which works with binned distributions (histograms). We denote $\mathbf{D} = (D_j)_{j=1, \dots, n_\mathrm{obs}}$ the vector of observed counts; an estimator of the true distribution $\overline{\mathbf{N}}$ will be denoted as $\hat{\mathbf{N}}= (\hat{N}_i)_{i=1, \dots, n_\mathrm{true}}$.
 
The unfolding matrix $\mathbf{U} = (U_{ij})$ is built as the transition matrix 
\begin{equation}
   U_{ij} \equiv  \mathbb{P}(X^\mathrm{true}~ \mathrm{in~bin}~i \; | \; Y^\mathrm{obs}~\mathrm{in~bin}~j) \equiv  \mathbb{P}(X^\mathrm{true}_i \; | \; Y^\mathrm{obs}_j).
\end{equation} 
which can be written, using Bayes' formula, as 
\begin{equation}
    U_{ij} \equiv  \frac{\mathbb{P}(Y^\mathrm{obs}_j \; | \; X^\mathrm{true}_i) \; \mathbb{P}(X^\mathrm{true}_i) }{\mathbb{P}( Y^\mathrm{obs}_j)}.
\end{equation}
The denominator can be regarded as a normalization factor, ensuring that $\sum_i U_{ij} = 1$: all observed counts originate from some true bin. The reverted conditionnal probability $\mathbb{P}(Y^\mathrm{obs}_j \; | \; X^\mathrm{true}_i) $ is to be identified with the detector response matrix $\mathbf{R}$, transforming the true variable into the observed one. Finally, $\mathbb{P}(X^\mathrm{true}_i)$ is a prior guess of what the distribution of $X^\mathrm{true}$ could be, called in short \textit{prior} (denoted $\mathbf{P}_0$). The unfolding matrix then writes \begin{equation}
    U_{ij} = \frac{R_{ij} P_{0,i} }{\sum_{l} R_{lj} P_{0,l}},
\end{equation} built upon only two ingredients: the detector response matrix and the prior. The unfolded estimator $\hat{\mathbf{N}}$ is obtained as $\hat{\mathbf{N}} = \mathbf{U} \cdot \mathbf{D}$ as a direct consequence of the relation $\mathbb{P}(X^\mathrm{true}_i ) = \sum_j \mathbb{P}(X^\mathrm{true}_i \; | \; Y^\mathrm{obs}_j) \, \mathbb{P}(Y^\mathrm{obs}_j)$.

To mitigate the arbitrariness due to the choice of a specific prior, iterations are introduced. The prior is replaced, for the next iteration, by the true spectrum $\hat{\mathbf{N}}$ just extracted: $P_{1,i} \equiv \hat{N}_i / \sum_l \hat{N}_l$. The algorithm then reads:\begin{enumerate}
    \item Initialization: pick a prior $\mathbf{P}_0$;
    \item Recursion: for any iteration $k$\begin{itemize}
        \item  build the unfolding matrix $\mathbf{U}_k$ as \begin{equation}\label{eqn:unfold-matrix-k}
        U_{k,ij} = \frac{R_{ij} P_{k,i} }{\sum_{l} R_{lj} P_{k,l}} \,;
        \end{equation}
        \item extract the unfolded distribution $\hat{\mathbf{N}}_{k+1} = \mathbf{U}_k \cdot \mathbf{D}$;
        \item update the prior as $P_{k+1,i} \equiv \hat{N}_{k+1,i} / \sum_l \hat{N}_{k+1,l}$.
    \end{itemize}
\end{enumerate}
\noindent This algorithm produces a sequence of true spectra $(\hat{\mathbf{N}}_k)_{k \geqslant 1}$, for which the prior is (up to a normalization) nothing but the initial condition. The response matrix stays the same throughout all iterations and defines the endpoint of the sequence, that we may note $\hat{\mathbf{N}}_\infty$.

What are the properties of $\hat{\mathbf{N}}_\infty$? Few theoretical studies have been done on this iterative algorithm, through its connection to the expectation-maximization (EM) algorithm. Indeed, applying an EM algorithm on Poisson likelihoods leads to the exact same iteration (see a derivation in \cite[section 4.1.2]{KuuselaMSc}). Results of our interest here are as follows \cite[and references therein]{KuuselaMSc,KuuselaPhD}:\begin{enumerate}
    \item $\hat{\mathbf{N}}_\infty$ is a maximum-likelihood estimator (MLE) for the Poisson likelihood built from the observations $\mathbf{D}$:\begin{equation}
        \mathcal{L}(\mathbf{N}~;~ \mathbf{D}) = \prod_{j=1}^{n_\mathrm{obs}} \mathrm{Poisson}\big(D_j ~ ; ~ (\mathbf{RN})_j\big)~;
    \end{equation}
    \item it does not depends on the chosen prior;
    \item if the response matrix is perfectly known\footnote{The importance of the response matrix can be illustrated as follows in a simple situation. Let $\overline{\mathbf{R}}$ be the exact response matrix; observations are such that $\mathbf{D} \sim \mathrm{Poisson} (\overline{\mathbf{R}} \, \overline{\mathbf{N}})$. Assuming $\mathbf{R}$ to be an invertible square matrix, we may formally identify $\hat{\mathbf{N}}_\infty$ to $\mathbf{R}^{-1}\mathbf{D}$. As a result, $\langle \hat{\mathbf{N}}_\infty \rangle = \mathbf{R}^{-1} \,\overline{\mathbf{R}} \, \overline{\mathbf{N}} $ : the endpoint is biased when the reponse matrix in the unfolding is not the exact one ($\mathbf{R}\neq\overline{\mathbf{R}}$).} then $\hat{\mathbf{N}}_\infty$ is an unbiased MLE, \textit{i.e.} over statistical realizations we have $\langle \hat{\mathbf{N}}_\infty \rangle = \overline{\mathbf{N}}$ where $\overline{\mathbf{N}}$ is the true distribution.  
\end{enumerate}

Note that these properties hold under the assumption that the response has full column rank, \textit{i.e.} $\mathrm{rank}(\mathbf{R}) = n_\mathrm{true}$; if needed, the number of bins $n_\mathrm{true}$ of the unfolded distribution can be reduced until this condition is fulfilled. 

The three properties above are lost\footnote{A singular exception arises when the prior equals the true distribution ($\mathbf{P}_0 = \overline{\mathbf{N}}$), but never occurs in real data analysis where the true distribution is unknown.} when the algorithm is truncated after a finite number of iterations $k$. $\hat{\mathbf{N}}_k$ is not a maximum-likelihood estimator. It varies under a change of prior, and the lower $k$, the larger this variation. A truncation bias is introduced: even with the exact response matrix, we have $\langle \hat{\mathbf{N}}_k \rangle \neq \overline{\mathbf{N}}$, resulting in (severe) undercoverage. However, $\hat{\mathbf{N}}_\infty$ often suffers from large variance and lack of smoothness, which is a typical feature of anti-smearing processes. Therefore, limited deviations from these properties may be acceptable, but the choice of $k$ should be addressed, in any case, with special care; small values of k could lead to sizeable bias. We present in this paper a data-driven criterion to choose a suitable $k$ (cf. section~\ref{scn:criterion}). \\

Because of the iterative nature of the algorithm, uncertainty propagation is not straightforward. There are no linear relation between the unfolded  ($\hat{\mathbf{N}}_k$) and observed ($\mathbf{D}$) distributions for $k\geqslant 2$ since updated priors and unfolding matrices are also dependent on $\mathbf{D}$. However, an error propagation matrix $\mathbf{E}_k$ can be iteratively built \cite{adye2011} and allows to obtain the covariance matrix $\mathbf{V}_{k}$ of the unfolded spectrum analytically as \begin{equation}
    \mathbf{V}_{k} = \mathbf{E}_k \mathbf{V}_{\mathbf{D}}\, \mathbf{E}_k^T
\end{equation} where $\mathbf{V}_{\mathbf{D}}$ is the covariance matrix associated to the observed distribution $\mathbf{D}$. Another option is to numerically sample the covariance matrix $\mathbf{V}_{k}$ (this procedure is known as \textit{bootstrap resampling} in statistics), as follows: \begin{enumerate}
    \item build a set of toy spectra $\{\mathbf{D}^{(t)}\}_{t=1,\dots,N}$ following the variance $\mathbf{V}_{\mathbf{D}}$;
    \item unfold each toy spectrum separately to get $\{\hat{\mathbf{N}}_k^{(t)} \}_{t=1,\dots,N}$;
    \item an estimator of the variance $\mathbf{V}_{k}$ is given by \begin{equation} \label{eqn:sampled-cov}
       \hat{ \mathbf{V}}_{k} = \frac{1}{N-1} \sum_{t} \delta \hat{\mathbf{N}}_k^{(t)T} \cdot \delta \hat{\mathbf{N}}_k^{(t)}
    \end{equation} where $\delta \hat{\mathbf{N}}_k^{(t)} = \hat{\mathbf{N}}_k^{(t)} - \langle\hat{\mathbf{N}}_k^{(t)} \rangle$.
\end{enumerate}

Another source of variance of the unfolded spectrum is systematic and comes from uncertainties on the detector response matrix. Because the response matrix defines the endpoint $\hat{\mathbf{N}}_\infty$, a biased response would lead to a biased endpoint spectrum; it is illustrated in section~\ref{scn:wrong-response}. A first simple solution is to increase the covariance of the observed distribution $\mathbf{V}_{\mathbf{D}} \rightarrow \mathbf{V}_{\mathbf{D}} + \mathbf{V}_{\mathrm{syst}}$ to include systematic uncertainties, and propagate this new error matrix to the unfolded space using one of the methods described above (analytical or numerical). The systematic variance can also be evaluated from alternative unfoldings, built using modified response matrices in eqn.~(\ref{eqn:unfold-matrix-k}). From a set of response matrices $\{\mathbf{R}^{(r)} \}$ representative of the expected variations of $\mathbf{R}$, one could obtain a set of unfolded spectra $ \{\hat{\mathbf{N}}_k^{(r)} \}$. Their distribution allows to built a systematic covariance matrix for $\hat{\mathbf{N}}_k$ (and $\hat{\mathbf{N}}_\infty$) as in eqn.~(\ref{eqn:sampled-cov}). Whatever method is used, an useful validation is to check that the total variance $\mathbf{V}_k$ provides proper coverage for systematically and statistically fluctuating realizations. \\

So far, the observed data $\mathbf{D}$ has been assumed to be background-free, which is not a realistic case for most high-energy experiments. Backgrounds can be accounted for in several ways.\begin{enumerate}
    \item \textit{Background subtraction.} The MC background prediction $\mathbf{B}_\mathrm{MC}$ is subtracted from the observed data $\mathbf{D}$ and the unfolding is applied on the signal distribution $\mathbf{D} - \mathbf{B}_\mathrm{MC}$. This is however only relevant when the background prediction is known to be accurate.
    \item \textit{Scaling factor from control regions.} A common way to monitor the MC background prediction is to use control regions (sidebands, SB). The observed data/MC ratio $\alpha = N_\mathrm{Data}^\mathrm{SB} / N_\mathrm{MC}^\mathrm{SB} $ in the sideband is used to scale the background prediction in the region of interest; the unfolding is then applied on $\mathbf{D} - \alpha \,  \mathbf{B}_\mathrm{MC}$. To be used, this method requires: 1) to build signal-free sidebands and 2) that extrapolation of a single normalization-like factor $\alpha$ from sideband to the main sample is meaningful. The later is achieved when the background distribution in the control region closely relates to background in the main sample, e.g. if they share the same kinematic distribution, or type of interactions, etc.
    \item \textit{Simultaneous unfolding of signal and sidebands}. When signal events are observed in the sideband, the unfolding matrix can be extended to \begin{equation}
        \begin{pmatrix} \hat{\mathbf{N}}_\mathrm{Signal} \\ \hat{\mathbf{N}}_\mathrm{Bkgd}
        \end{pmatrix} = \mathbf{U} \cdot \begin{pmatrix} \mathbf{D} \\ \mathbf{D}_\mathrm{SB} \end{pmatrix}
    \end{equation} where $\mathbf{D}$ ($\mathbf{D}_\mathrm{SB}$) is the observed distribution of events in the main sample (sideband), and $\hat{\mathbf{N}}_\mathrm{Signal}$ ($\hat{\mathbf{N}}_\mathrm{Bkgd}$) the unfolded distribution of signal (background) events. This method is also useful when data/MC shape discrepancies are observed in background distributions, making the norm correction of method 2 inappropriate.
\end{enumerate}
On a statistical point of view, an asset of the last method is to preserve the Poisson properties of the input distribution $(\mathbf{D}, \mathbf{D}_\mathrm{SB})$. In (scaled) background subtraction, it is therefore not guaranteed that properties of the endpoint $\hat{\mathbf{N}}_\infty$ are preserved, in particular that it provides an unbiased MLE.

In the following we will assume the input distribution to follow Poisson statistics. Up to redefinition of binnings, simple or simultaneous unfoldings are equivalent, and need not to be treated differently.

\section{The 2-peak model}\label{scn:2peak-model}
\begin{figure}[!ht]
	\begin{center} \includegraphics[width=\linewidth]{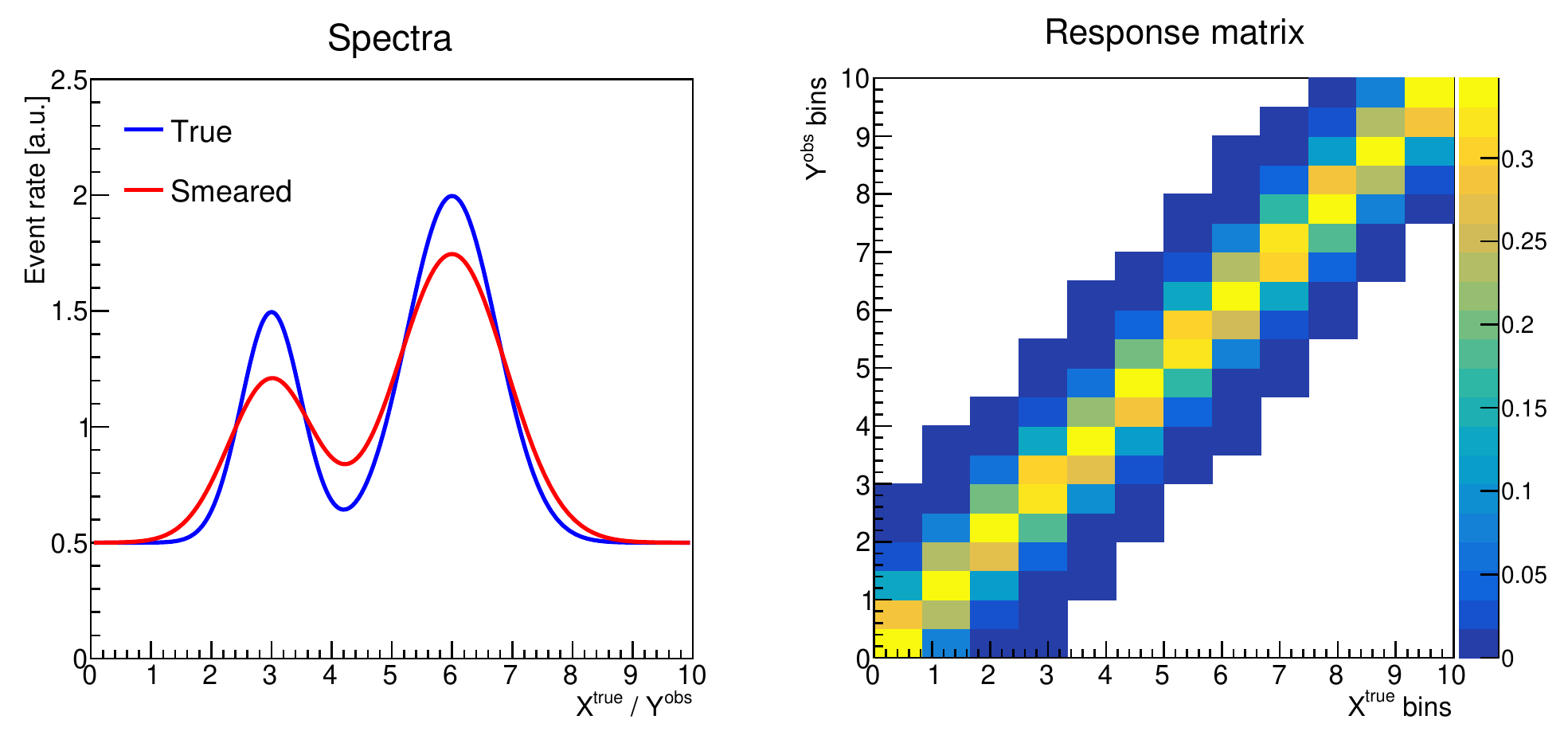} \\
 (a)  \hspace{6.5cm} (b)
\end{center}
	\caption{(a) True and smeared spectra for the nominal two-peak model. (b) The corresponding response matrix, with $n_\mathrm{true}=12$ and $n_\mathrm{obs}=20$.}
	\label{fig:2peak-model}
\end{figure}

Throughout this paper we will use a simple two-peak model to illustrate the behaviour of the iterative unfolding. The true two peaks distribution is smeared by an artificial detector smearing, defined here as a convolution by a gaussian of width $\sigma_s = 0.5$. The true spectrum and the smeared spectrum are displayed in figure~\ref{fig:2peak-model} along with the response matrix. The number of bins is set to $n_\mathrm{obs}=20$ for observed data and $n_\mathrm{true}=12$ for the unfolded spectrum. The parameters of the two peaks are given in table~\ref{tab:models-def}. Modified two-peak models, used to test the properties of the algorithm (cf. sections~\ref{scn:DD-criterion}-\ref{scn:wrong-response}), are also introduced in table~\ref{tab:models-def}.

\begin{table}[!h]
\centering
   \begin{tabular}{|c|l||c|c|c||c|}

    \multicolumn{2}{|c||}{Toy model} & First peak & Second peak & Baseline & Smearing \\
    \# & Name & $(A_1,\mu_1,\sigma_1)$ & $(A_2,\mu_2,\sigma_2)$ & $b$ & $\sigma_s$ \\
    \hline
 1 &  Nominal & (1, 3, 0.5) & (1.5, 6, 0.75) & \multirow{5}{*}{0.5} & \multirow{5}{*}{0.5}\\
 2 & Shifted peaks & (1, \textbf{3.2}, 0.5)  & (1.5, \textbf{6.2}, 0.75)&  & \\
 3 & Closer peaks & (1, \textbf{3.2}, 0.5) &  (1.5, 6, 0.75) & & \\ 
 4 & Wider 2nd peak & (1, 3, 0.5) & (1.5, 6, \textbf{0.9})&  & \\
  5 & Smaller 1st peak & (\textbf{0.75}, 3, 0.5) & (1.5, 6, 0.75)&  & \\ \hline
  $1n$ & More smearing & \multicolumn{3}{c||}{ Same as toy model \#$n$} & \textbf{0.55} \\
     \end{tabular}
     \caption{Definition of two-peak toy models used in this article. $(A_i,\mu_i,\sigma_i)$ refers to the maximum amplitude, mean position, width of the $i$-th peak. Differences with respect to the nominal model are highlighted.}
    \label{tab:models-def}
\end{table}

\section{Data-driven convergence criterion }\label{scn:DD-criterion}

\subsection{Notations}
\noindent Let us first summarize some notations and definitions used through this paper. We define the following vectors of size $n \equiv n_\mathrm{true}$ (number of bins in the phase-space of true variables): \begin{itemize}
   \item $\overline{\mathbf{N}}$ is the true spectrum, or \textit{truth};
    \item $\hat{\mathbf{N}}_k$ is the unfolded spectrum (true spectrum estimator) after $k$ iterations of the algorithm;
    \item $\langle \hat{\mathbf{N}}_k \rangle \equiv \mathbb{E}(\hat{\mathbf{N}}_k)$ is the expected value of the unfolded spectrum after $k$ iterations;
    \item $\mathbf{b}_k = \langle \hat{\mathbf{N}}_k \rangle  - \overline{\mathbf{N}}$ is the average bias.
\end{itemize}
We also denote the following $n \times n$ matrix:\begin{itemize}
    \item $\mathbf{V}_k \equiv \textrm{Var}(\hat{\mathbf{N}}_k) = \mathbb{E} \big[ \big(\hat{\mathbf{N}}_k - \langle \hat{\mathbf{N}}_k \rangle \big) \big(\hat{\mathbf{N}}_k - \langle \hat{\mathbf{N}}_k \rangle \big)^T  \big]$ is the covariance matrix associated to the unfolded spectrum $\hat{\mathbf{N}}_k$.
\end{itemize}
Finally, we also use the following metrics to study the convergence of the algorithm and establish the convergence criterion:\begin{itemize}
    \item a measure of the distance of the unfolded spectrum to the truth \begin{equation}
        \chi^2_\mathrm{true}(k) = \big(\hat{\mathbf{N}}_k -  \overline{\mathbf{N}} \big)^T \cdot \mathbf{V}_{k}^{-1} \cdot \big(\hat{\mathbf{N}}_k -  \overline{\mathbf{N}} \big);
    \end{equation}
    \item a measure of the distance of the unfolded spectrum to the endpoint \begin{equation}
        \chi^2_\mathrm{data}(k) = \big(\hat{\mathbf{N}}_k - \hat{\mathbf{N}}_\infty \big)^T \cdot \mathbf{V}_{k}^{-1} \cdot \big(\hat{\mathbf{N}}_k - \hat{\mathbf{N}}_\infty \big); 
    \end{equation} for which we have by construction $\lim_{k \to \infty} \chi^2_\text{data}(k) = 0 $;
    \item a measure to compare bias and variance: \begin{equation}
        \chi^2_\mathrm{bias}(k) = \mathbf{b}_{k}^T\, \mathbf{V}_{k}^{-1}\, \mathbf{b}_{k}.
    \end{equation}The following useful relation (proof in appendix) describes how the distribution of $\chi^2_\mathrm{true}$ departs from a perfect $\chi^2$ law in presence of bias:
    \begin{equation}\label{eqn:E(chi2true)}
        \mathbb{E} \big[ \chi^2_\mathrm{true}(k) \big] = n + \chi^2_\mathrm{bias}(k).
    \end{equation}
\end{itemize}

\subsection{Convergence criterion}\label{scn:criterion}

The iterative unfolding produces, from a given input data spectrum, a sequence of spectra with associated covariances $(\hat{\mathbf{N}}_{k}, \mathbf{V}_{k})_{k \geqslant 1}$. We would like to build a data-driven criterion, \textit{i.e.} to be applied only on this sequence of spectra and covariances and not on an \textit{a priori} MC distribution, to determine what is an appropriate number of iterations to unfold this particular input data spectrum.

Because the algorithm is truncated (finite number of iterations $k$) it is expected to have a convergence bias $\mathbf{b}_{k} \equiv \langle \hat{\mathbf{N}}_{k} \rangle  - \overline{\mathbf{N}} \neq \mathbf{0}$. For this bias to have limited impact on the coverage -- defined at this point by the covariance $\mathbf{V}_{k}$ -- we would like to keep it "well smaller than the error bars". In other words, we wish to have 
\begin{equation}\label{eqn:chi2-bias}
    \chi^2_\text{bias}(k) \equiv \mathbf{b}_{k}^T \mathbf{V}_{k}^{-1}  \mathbf{b}_{k}  \leqslant n\varepsilon^2
\end{equation} for some $\varepsilon^2$ much smaller than 1, or equivalently from eqn.~(\ref{eqn:E(chi2true)}) \begin{equation}
    \mathbb{E} \big[ \chi^2_\mathrm{true}(k) \big] \leqslant n ( 1+\varepsilon^2).
\end{equation} The impact of the size of bias $\varepsilon^2$ on the coverage is studied in section \ref{scn:coverage}.

This $\chi^2_\text{bias}$ is not computable for actual data since biases are unknown. Our proposal is to use the endpoint spectrum $\hat{\mathbf{N}}_{\infty}$ as a pivot. This endpoint is an unbiased\footnote{We insist once again that this only holds when the response matrix is exactly known. Other cases are discussed in section~\ref{scn:wrong-response}.} MLE of the true distribution (cf.~section~\ref{scn:presentation}), leading to $\mathbf{b}_\infty = 0$. Consequently, if the unfolded spectrum $\hat{\mathbf{N}}_{k}$ is close enough to the endpoint, we can expect it to be close to the true spectrum as well. Formally, we define the distance to the endpoint $\mathbf{d}_k =  \hat{\mathbf{N}}_k - \hat{\mathbf{N}}_\infty$  
and the $\chi^2_\text{data}$ metrics as \begin{equation}
    \chi^2_\text{data}(k) \equiv \mathbf{d}_{k}^T \mathbf{V}_{k}^{-1}  \mathbf{d}_{k}
\end{equation} which describes the level of convergence of the algorithm. As it is built only from the sequence of unfolded spectra $(\hat{\mathbf{N}}_{k}, \mathbf{V}_{k})_{k \geqslant 1}$, this quantity can be used to construct a data-driven convergence criterion. 

Using toy studies for which the truth is known, the goal is to find some $\eta^2$ such that \begin{equation}\label{eqn:criterion}
    \mathbb{E} \big[ \chi^2_\mathrm{data}(k) \big] \leqslant n \, \eta^2 \quad \Longrightarrow \quad \mathbb{E} \big[ \chi^2_\mathrm{true}(k) \big] \leqslant n\, (1+ \varepsilon^2).
\end{equation}
This relation describes an average behaviour, determined over many statistical fluctuations. Since we only have a single realization of the real experiment, the condition to be applied on the unfolded data is simply \begin{equation}\label{eqn:criterion-data}
   \chi^2_\mathrm{data}(k)  \leqslant n \, \eta^2.
   \end{equation}
The number of iterations $k_0$ chosen to truncate the algorithm will be the smallest $k$ satisfying the above relation.

Typical MC-driven convergence criteria would only rely on conditions similar to the right-hand side of eqn.~(\ref{eqn:criterion}). Looking at the evolution of biases using fake-data sets would provide the number of iterations $k_0$, usually very small ($k_0<5$). The value of $k_0$ and the corresponding spectrum $\hat{\mathbf{N}}_{k_0}$ obtained this way are largely correlated to the choice of prior, and the amount of bias after only a few iterations depends on how different \textcolor{red}{is} the truth from the prior. In particular, the (unknown) true distribution of real data may be further from the prior that what has been tested with toy models. The data-driven criterion of eqn.~(\ref{eqn:criterion-data}) relies instead on the endpoint spectrum, which does not vary with the chosen prior. Whenever the difference truth/prior is higher on real data, the endpoint spectrum remains a robust quantity upon which a convergence criterion may be built.

Admittedly, the value of $\eta^2$ in eqn.~(\ref{eqn:criterion-data}) is chosen using MC toy models, and this criterion is not fully data-driven. However, the convergence speed of $\chi^2_\mathrm{data}$ is characterized by the response matrix used in the unfolding, which introduce much less model-dependence than the choice of a specific prior. Even if $\eta^2$ is chosen on some toy models, the criterion (\ref{eqn:criterion-data}) is therefore still relevant for real data, where the truth is unknown.

\subsection{Coverage evolution and value of $\varepsilon^2$}\label{scn:coverage}
Let us study now how the coverage provided by the covariance $\mathbf{V}_k$ evolves with the number of iterations $k$, and investigate its relation with the measure of bias $\chi^2_\mathrm{bias}$. For a given $k$ and a given realization of the experiment $\hat{\mathbf{N}}_k$, the \textit{confidence region} for $1-\alpha$ CL is defined as the set of spectra $\mathbf{N}$ such that \begin{equation}\label{eqn:conf-region}
\chi^2_k (\mathbf{N}) \leqslant \chi^2_\text{lim} (\alpha),
\end{equation} with \begin{equation}
    \chi^2_k (\mathbf{N}) = \big( \hat{\mathbf{N}}_k - \mathbf{N} \big)^T \mathbf{V}^{-1}_k \big( \hat{\mathbf{N}}_k- \mathbf{N}  \big)
\end{equation}
and \begin{equation}\label{eqn:conf-region-size}
    \chi^2_\text{lim}(\alpha) = F_n(1-\alpha)
\end{equation} where $F_n$ is the inverse of the cumulative function of a $\chi^2$ distribution with $n$ degrees of freedom \cite{PDGstatistics} (in the large sample approximation). The \textit{coverage} of this confidence region is defined as the fraction of statistical realizations $\hat{\mathbf{N}}_k$ for which the true spectrum $\overline{\mathbf{N}}$ belongs to the confidence region, that is \begin{equation}
    \chi^2_k (\overline{\mathbf{N}}) \equiv \chi^2_\mathrm{true}(k) \leqslant \chi^2_\text{lim}(\alpha).
\end{equation}

\begin{figure}[!ht]
	\begin{center}
\begin{tabular}{cc}
   \includegraphics[width=0.49\linewidth]{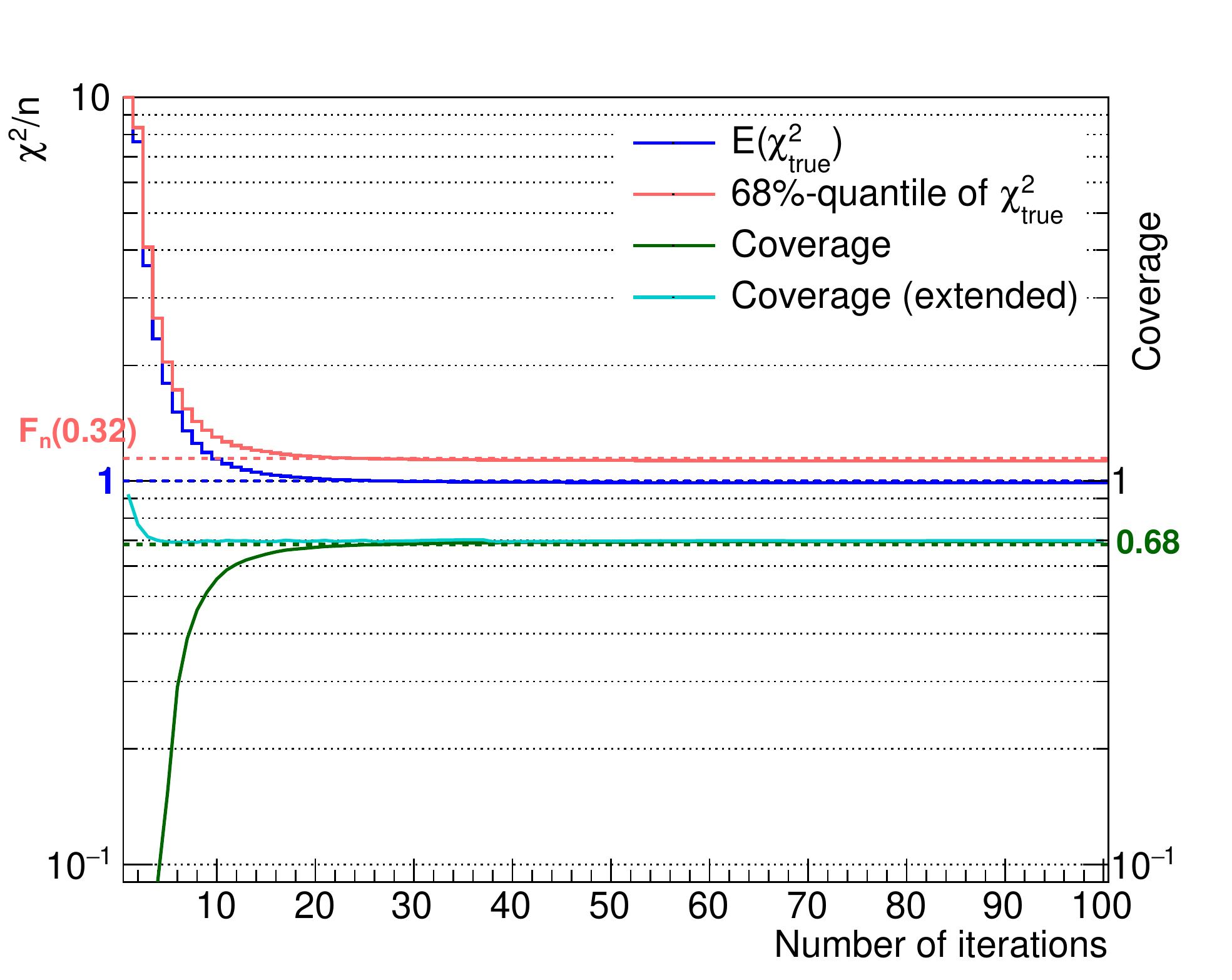}  &  
   \includegraphics[width=0.49\linewidth]{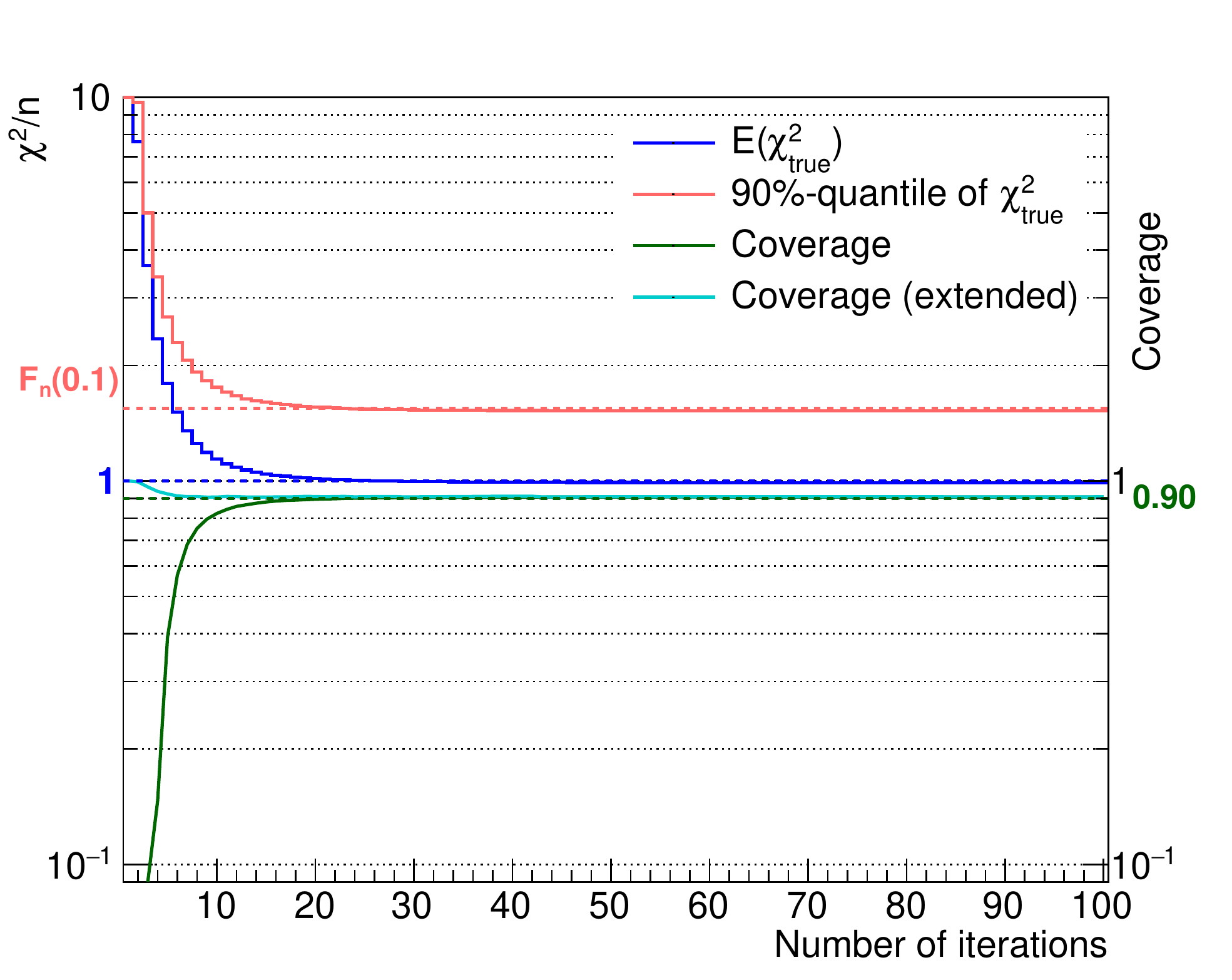} \\
    (a) & (b) 
\end{tabular}
\end{center}
	\caption{Evolution of coverage of confidence regions for (a) 68\% CL and (b) 90\% CL, produced using model \#3. The exact response matrix is used, leading to $\lim_{k \to \infty} \mathbb{E}[\chi^2_\mathrm{true}(k)]/n=1$ (unbiased endpoint). The 68\%-quantile of $\chi^2_\mathrm{true}$ is the value of $\chi^2_\text{lim}$ that provides a 68\% CL confidence region with proper coverage; it corresponds to $F_n (1-\alpha)$ (red dotted line) in the no-bias limit (large $k$). The extended confidence region defined by $\chi^2_\text{lim}(\alpha; \chi^2_\mathrm{bias})$ achieves correct coverage even in presence of large biases.}
	\label{fig:coverage}
\end{figure}

When there is no bias ($\chi^2_\mathrm{bias}=0$), the region defined by eqns.~(\ref{eqn:conf-region}-\ref{eqn:conf-region-size}) achieves the nominal $1-\alpha$ coverage; however, in presence of bias, the coverage is reduced. This is illustrated in figure~\ref{fig:coverage} using toy model \#3 (see definition in table~\ref{tab:models-def}). The exact response matrix is used, leading to $\lim_{k \to \infty} \mathbb{E}[\chi^2_\mathrm{true}(k)]=n$ or equivalently $\lim_{k \to \infty} \chi^2_\mathrm{bias}(k)=0$. Large biases are observed for $k\lesssim 10$, resulting in poor coverage. With more iterations ($10 \lesssim k \lesssim 20$), biases reduce and coverage improves:  $>60\%$ (resp. $>80\%$) for the nominal 68\% CL (resp. 90\% CL) confidence region. For large number of iterations ($k\gtrsim 20$ in this example) there is almost no bias and the coverage is as expected.

The data-driven criterion~(\ref{eqn:criterion}) is built to provide a number of iterations $k_0$ allowing for a control of the truncation bias: $\chi^2_\mathrm{bias}(k_0) \leqslant n\varepsilon^2$. Nonetheless, the confidence region defined with the nominal $\chi^2_\text{lim}(\alpha) = F_n(1-\alpha)$ undercovers in presence of even small bias; one could then seek to increase this boundary to recover an appropriate coverage. We found that \begin{equation}\label{eqn:confidence-region-size-ext}
     \chi^2_\text{lim}\big(\alpha; \chi^2_\mathrm{bias}\big) \equiv (1 + \chi^2_\mathrm{bias}/n) \cdot F_n (1-\alpha)
\end{equation} is a good approximation, valid for a wide range of values of $\chi^2_\mathrm{bias}$ and $\alpha$. The coverage provided by eqn.~(\ref{eqn:confidence-region-size-ext}) is illustrated in figure~\ref{fig:coverage} for $1-\alpha$ being 68\% or 90\%; it is very satisfactory for $\chi^2_\mathrm{bias}/n \lesssim 1$. On real data analysis where $\chi^2_\mathrm{bias}(k_0)$ is unknown, a conservative confidence region can be defined using
\begin{equation}\label{eqn:confidence-region-size-ext2}
    \chi^2_\text{lim}(\alpha; n\varepsilon^2) = (1 + \varepsilon^2) \cdot F_n (1-\alpha).
\end{equation}

We have illustrated in this section how the coverage evolves in presence of bias, and how confidence regions can be extended to compensate the undercoverage induced by such biases. The choice of $\varepsilon^2$ in the criterion~(\ref{eqn:criterion}) is left to the discretion of the analyzer; but some recommendations follow. \begin{enumerate}
    \item For $\varepsilon^2 \ll 1$, the undercoverage is negligible and the unfolded spectrum $\hat{\mathbf{N}}_{k_0}$ and its covariance $\mathbf{V}_{k_0}$ can be used to define confidence regions. However, note that when the response matrix is not accurately known we have $\lim_{k \to \infty} \chi^2_\mathrm{bias}(k)/n = b_\infty > 0$ so arbitrarily small $\varepsilon^2$ are not possible (such cases are discussed in section~\ref{scn:wrong-response}).
    \item For $\varepsilon^2 < 1$, the loss of coverage may become significant and should be accounted for. Conservative confidence regions providing \textit{at least} nominal coverage can be recovered using $\chi^2_\text{lim}(\alpha; n\varepsilon^2)$ from eqn.~(\ref{eqn:confidence-region-size-ext2}), or equivalently by inflating the covariance matrix as $\mathbf{V}_{k_0} \to \mathbf{V}_{k_0}(1+\varepsilon^2)$.
\end{enumerate}

\subsection{Data-driven criterion: illustration}\label{scn:DD-illustration}
Having set the value of $\varepsilon^2$, the remaining task is to find $\eta^2$ fulfilling the condition \begin{equation}\label{eqn:criterion2}
    \mathbb{E} \big[ \chi^2_\mathrm{data}(k) \big] \leqslant n \, \eta^2 \quad \Longrightarrow \quad \mathbb{E} \big[ \chi^2_\mathrm{true}(k) \big] \leqslant n\, (1+ \varepsilon^2).
\end{equation}
 This section illustrates how $\eta^2$ can be chosen. Since the amount of bias and its evolution over iterations depend on the unknown true model, extracting $\eta^2$ from the nominal model (or standard prediction, or MC prediction) is not enough. It is important to evaluate what alternative true models could be plausible. For the two-peak model, we considered (cf. table~\ref{tab:models-def}): both peak being shifted in the same direction (model \#2); closer peaks (\#3); one peak being wider (\#4); and one peak being smaller (\#5). For now, the smearing is considered to be accurately known.

\begin{figure}[!ht]
	\begin{center}
\begin{tabular}{cc}
   \includegraphics[width=0.49\linewidth]{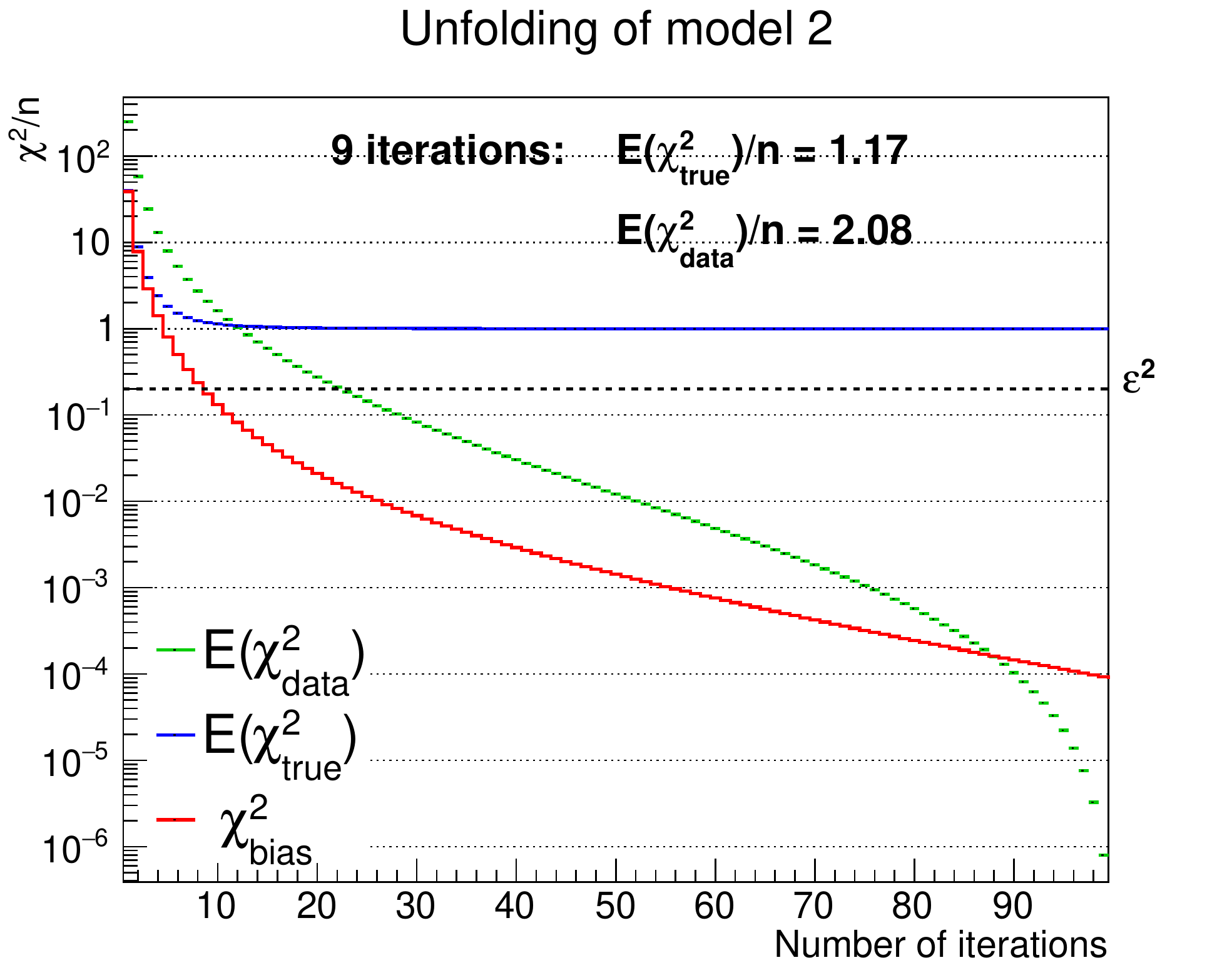}  &
   \includegraphics[width=0.49\linewidth]{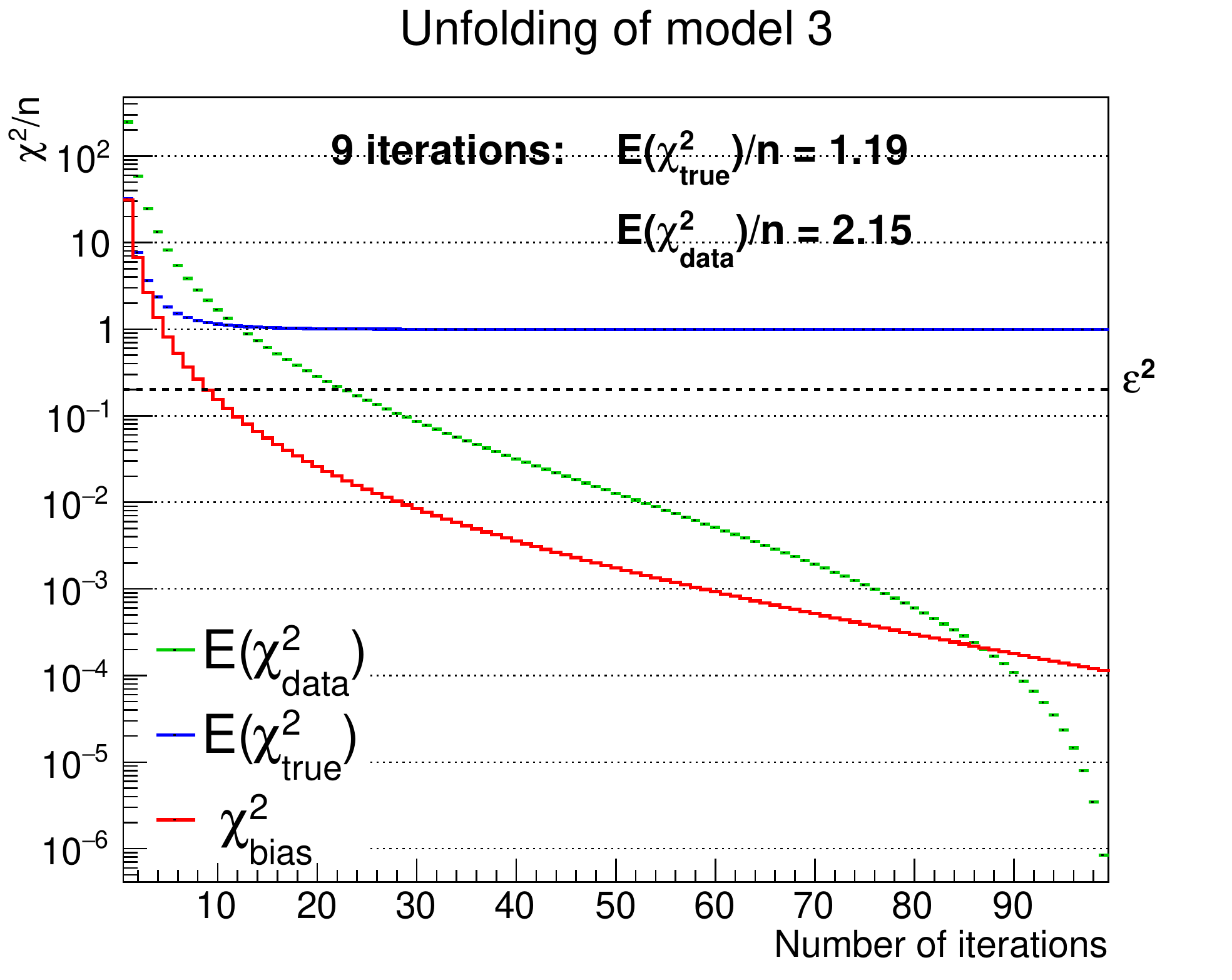} \\
   \includegraphics[width=0.49\linewidth]{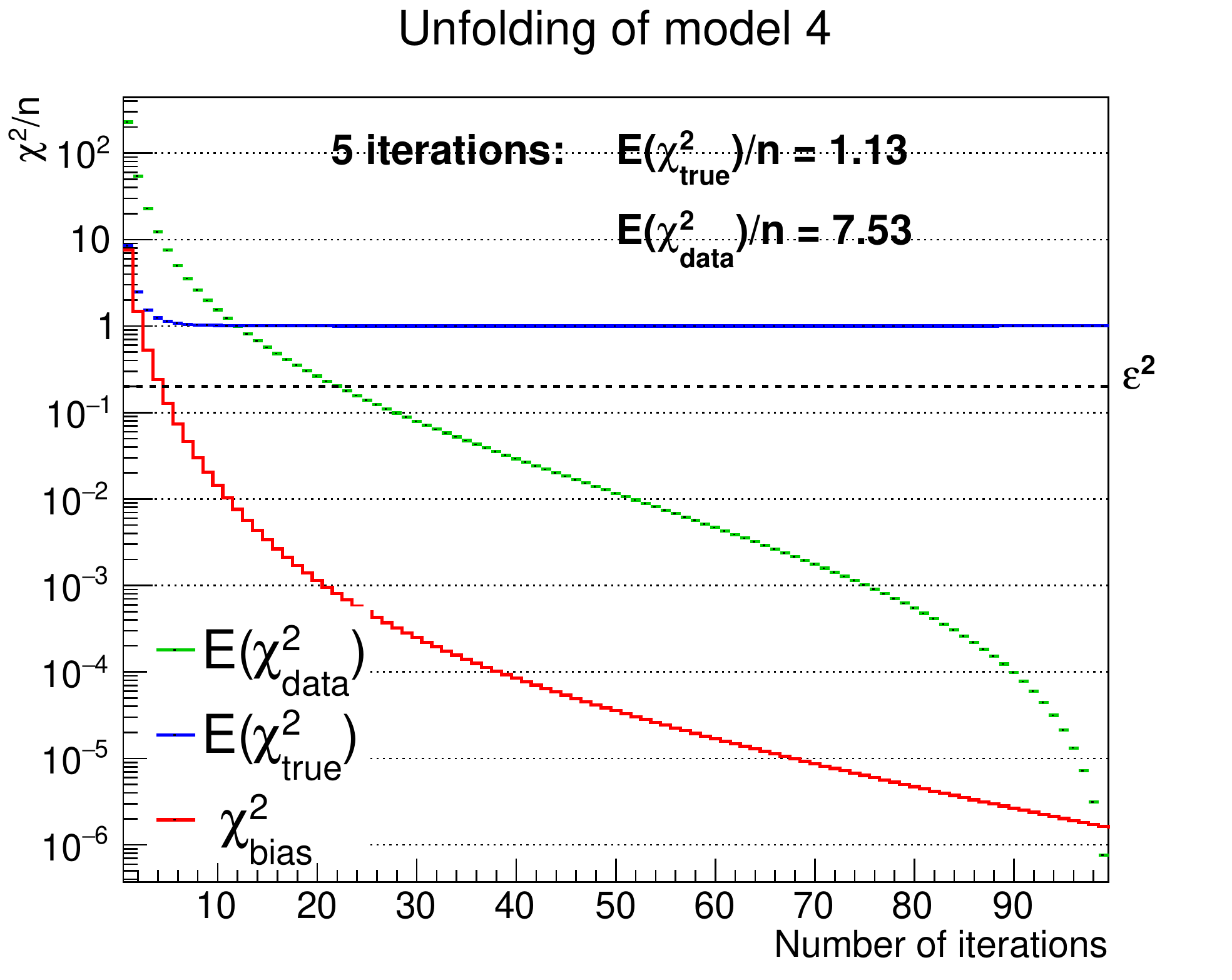}  &
   \includegraphics[width=0.49\linewidth]{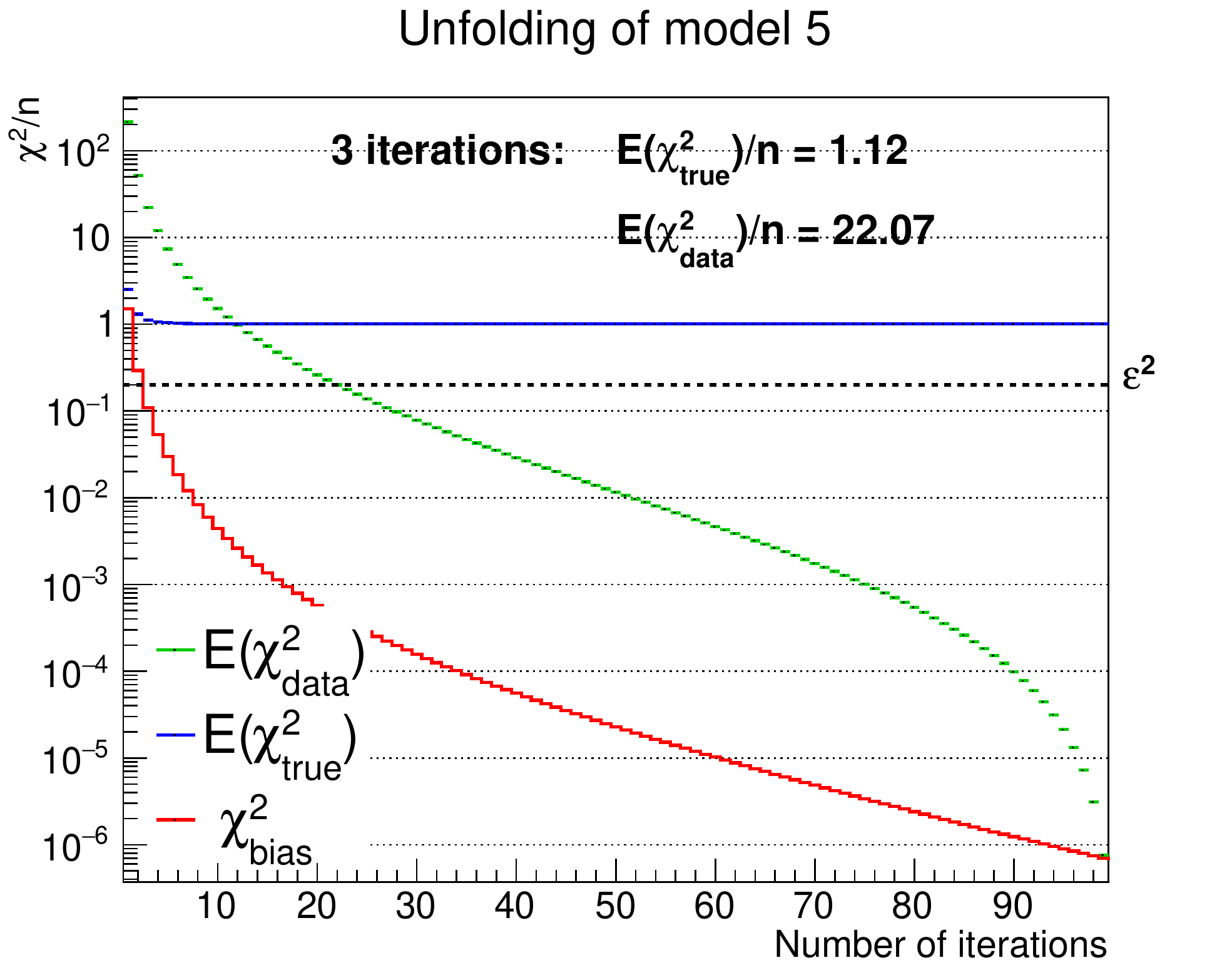}  \\
\end{tabular}
\end{center}
	\caption{Evolution of $\mathbb{E}(\chi^2_\mathrm{data}/n)$ and $\chi^2_\mathrm{bias}/n = \mathbb{E}( \chi^2_\mathrm{true}/n)-1$ for toy models \#2-5. For each model, the true response is used in the unfolding, yielding vanishing biases in the limit $k\to \infty$. The first iteration such that $\mathbb{E}(\chi^2_\mathrm{true}/n) \leqslant 1+\varepsilon^2$ is indicated, with $\varepsilon^2 = 0.2$. The corresponding values of $\mathbb{E}(\chi^2_\mathrm{data}/n)$ are indicated as well; $\eta^2$ is set as the lowest $\mathbb{E}(\chi^2_\mathrm{data}/n)$ among all models.}
	\label{fig:chi2-data-trueR}
\end{figure}

The combined evolution of $\mathbb{E}(\chi^2_\mathrm{data})$ and $\mathbb{E}( \chi^2_\mathrm{true})$ for models \#2-5 is shown in figure~\ref{fig:chi2-data-trueR}. The value of $\varepsilon^2$ is set here to $0.2$. For each model, the true response is used in the unfolding, but the prior is based on the nominal model (\#1). The number of iterations required to reach $\mathbb{E}( \chi^2_\mathrm{true}) \leqslant n(1+\varepsilon^2)$ -- or equivalently $\chi^2_\mathrm{bias} \leqslant \varepsilon^2$ --  varies from 3 to 9. When compared to the nominal model, largest shape discrepancies occur for models with shifted peaks (\#2-3): the value of $\chi^2_\mathrm{bias}(k=1)$, correlated to the difference prior/truth, is then the largest for these models. In turn, more iterations are required to reach below the $\varepsilon^2$ threshold.

\begin{figure}[!ht]
	\begin{center}
   \begin{tabular}{cc}
   \includegraphics[width=0.49\linewidth]{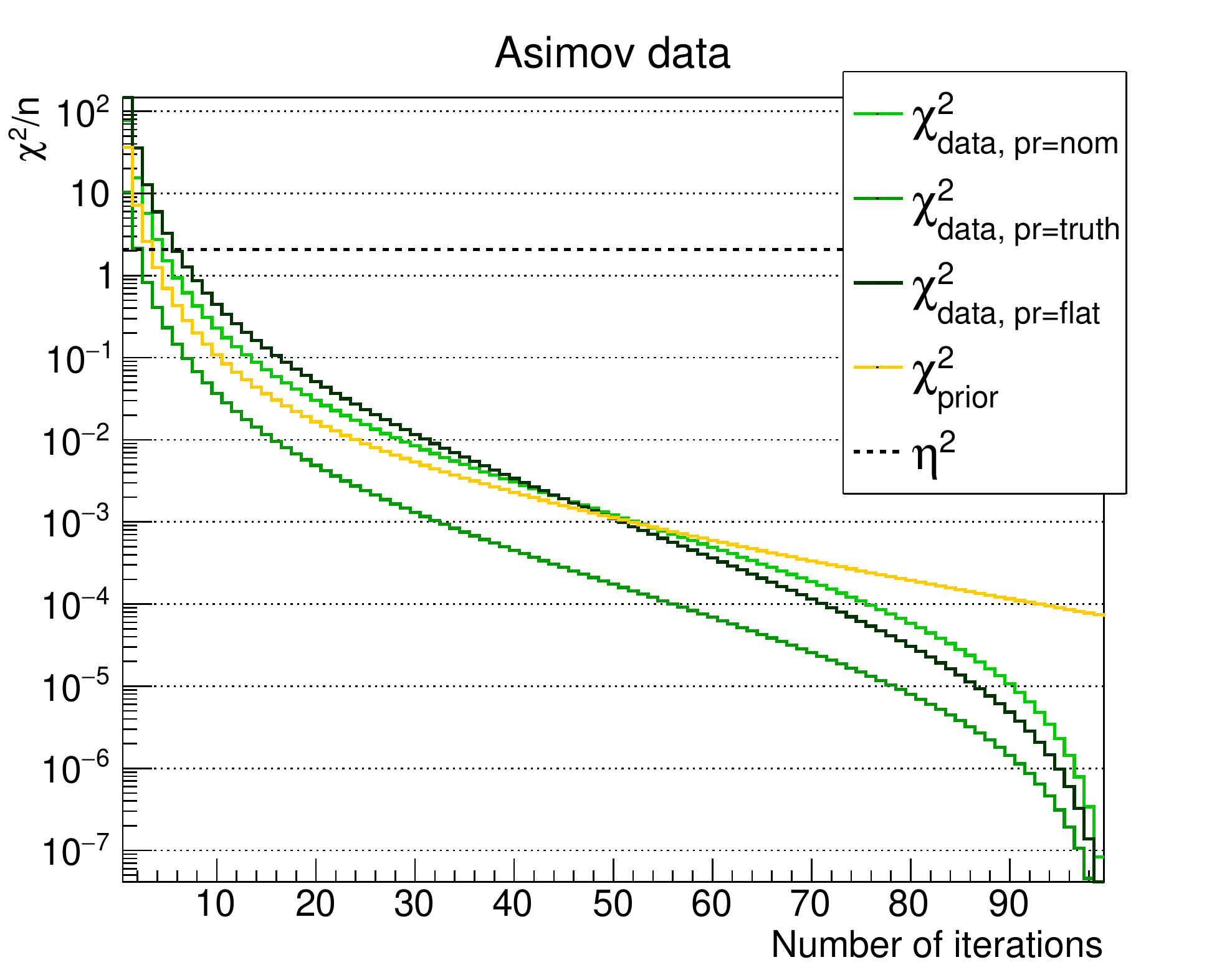}  &
   \includegraphics[width=0.49\linewidth]{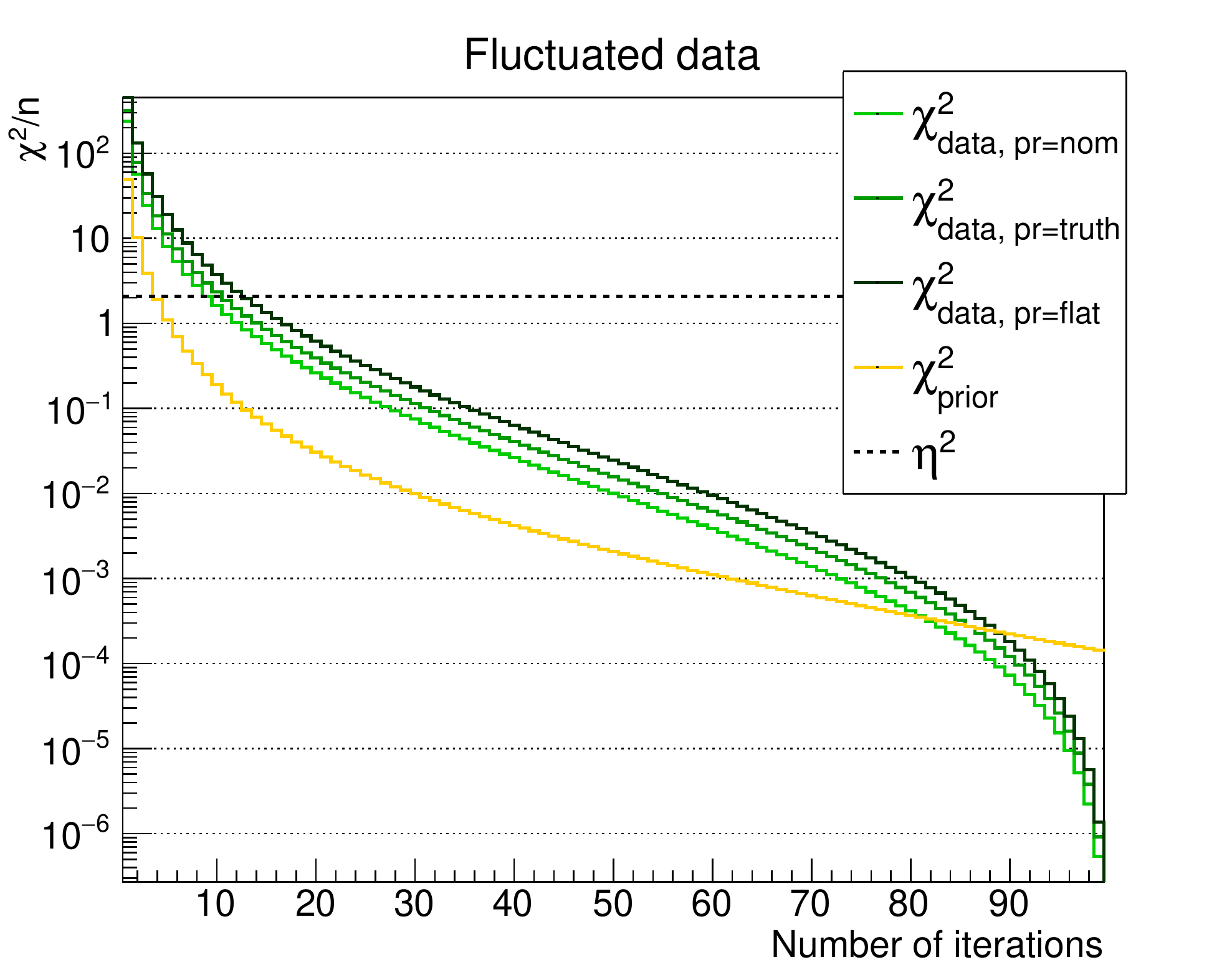} \\ (a) & (b)
\end{tabular} \end{center}
    	\caption{Evolution of a $\chi^2_\mathrm{data}$ for a single realization from model \#2, using several priors (nominal=model \#1, truth=model \#2, or flat). Depending on the chosen prior, the number of iterations given by the criterion of eqn.~(\ref{eqn:criterion-single}) is (a) 5, 2, or 6 for the Asimov data, and (b) 9, 11 or 13 for the random fluctuation.}
	\label{fig:chi2-data-singletoy}
\end{figure}

The values of $\mathbb{E}(\chi^2_\mathrm{data}/n)$ range from 2.08 to 22.1. By chosing $\eta^2$ as the lowest $\mathbb{E}(\chi^2_\mathrm{data}/n)$ among all models (in our example $\eta^2 = 2.08$), we ensure that for all models the condition (\ref{eqn:criterion2}) is satisfied. On figure~\ref{fig:chi2-data-singletoy}, the criterion \begin{equation}\label{eqn:criterion-single}
    \chi^2_\mathrm{data}(k) \leqslant n \, \eta^2
\end{equation} is applied on single realizations of the experiment (assuming model \#2 is true): the Asimov data, and a random fluctuation. In both cases, one observes that the number of iterations given by the convergence criterion varies with the input prior: for Asimov (fluctuated) data, one gets $k_0=5$ (9) with model \#1 as prior; $k_0=2$ (11) with model \#2 as prior; $k_0=6$ (13) for a flat prior. We can expect $k_0$ to be smallest when the prior is the truth; this is verified with Asimov data. However, because of random statistical fluctuations, the initial data set may appear more similar to another model: for fluctuated data, the lowest $k_0$ is with model \#1 as prior. This illustrates that the number of iterations given by eqn.~(\ref{eqn:criterion-single}) adapts to the difference between prior and truth\footnote{It would have been closer to real conditions to set the prior and vary the true model. However, in order compare the behaviour on the same fluctuation, the prior has been varied for a fixed truth (and realization).}. 

We also displayed \begin{equation}
    \chi^2_\mathrm{prior}(k) = \Delta \hat{\mathbf{N}}_k ^T \, \mathbf{V}_k^{-1} \, \Delta \hat{\mathbf{N}}_k \; ,
\end{equation} with $\Delta \hat{\mathbf{N}}_k = \hat{\mathbf{N}}_k - \hat{\mathbf{N}}'_k$ the difference induced by changing the prior in the unfolding: $\hat{\mathbf{N}}_k$ is obtained using the nominal prior and $\hat{\mathbf{N}}'_k$ using the truth as prior. This emphasizes that the endpoint spectrum does not depend on the selected prior, \textit{i.e.} $\lim_{k\to\infty} \chi^2_\mathrm{prior}(k)=0$.

\section{Imperfectly known response matrix}\label{scn:wrong-response}
So far, the basic case of a perfectly known response matrix has been discussed. In full generality, the bias from the algorithm at iteration $k$ can be written as\begin{equation}
    \mathbf{b}_k \equiv \langle \hat{\mathbf{N}}_k \rangle - \overline{\mathbf{N}} = \underbrace{\langle \hat{\mathbf{N}}_k \rangle - \langle\hat{\mathbf{N}}_\infty \rangle}_\mathrm{truncation~bias} \;  +  \; \underbrace{\langle\hat{\mathbf{N}}_\infty \rangle - \overline{\mathbf{N}}}_\mathrm{endpoint~bias}
\end{equation}
with the truncation bias vanishing in the limit $k \to \infty$. As discussed in section~\ref{scn:presentation}, the endpoint spectrum $\hat{\mathbf{N}}_\infty$ is an unbiased MLE (i.e. $\langle \hat{\mathbf{N}}_\infty \rangle= \overline{\mathbf{N}}$) when the response matrix is exactly known, but there remains a non-zero bias $\mathbf{b}_\infty$ (called here \textit{endpoint bias}) otherwise. In fact, we claim that in most data analyses the response matrix is not perfectly accurate and endpoint biases should be considered. We consider three sources which can bias the response matrix:
\begin{itemize}
    \item imperfect or biased knowledge of the detector response.
    \item finite binning of the true distributions.
    \item limited statistics of the simulation which might be used to built the response matrix.
\end{itemize}

\begin{figure}[!ht]
	\begin{center}
\begin{tabular}{cc}
   \includegraphics[width=0.49\linewidth]{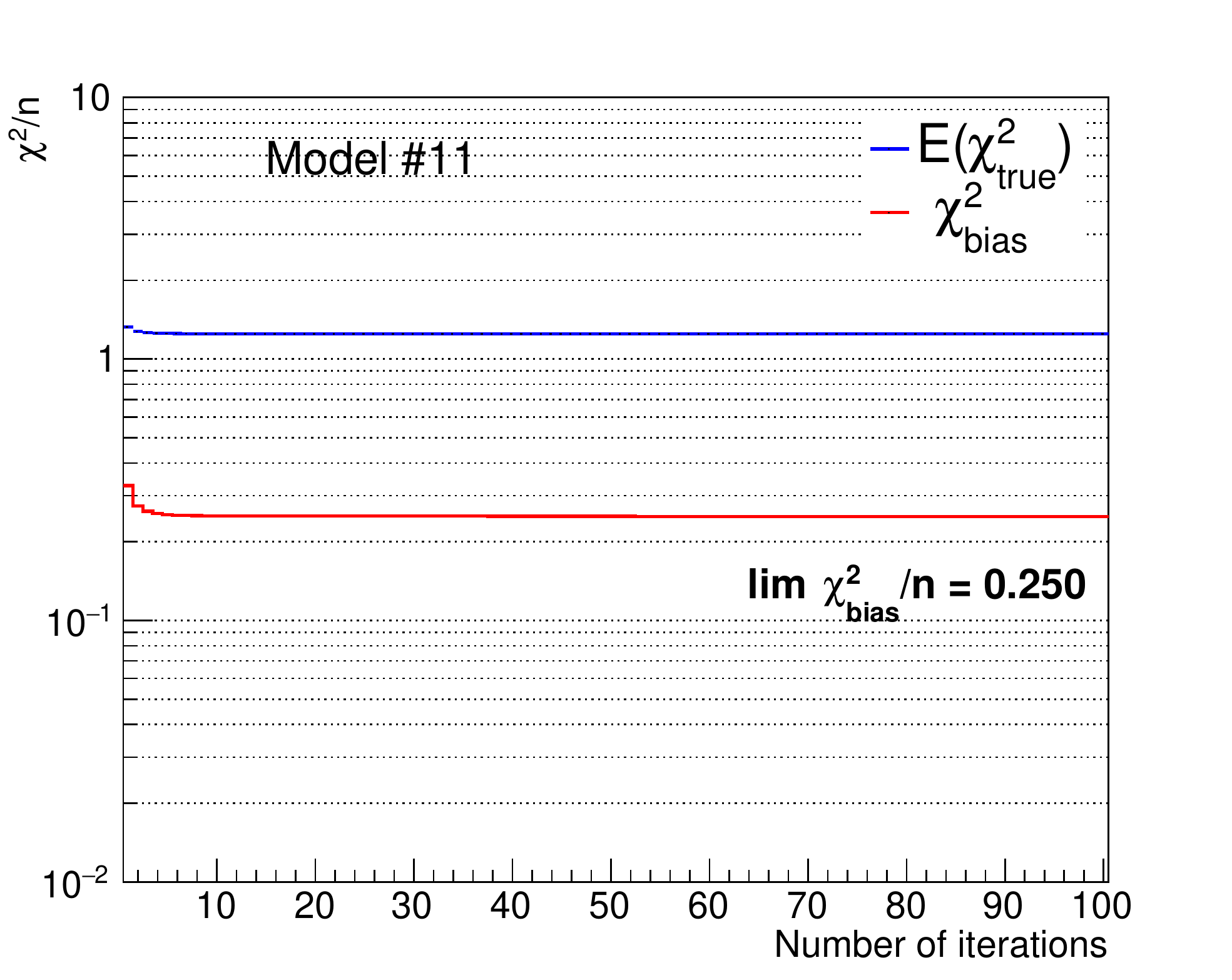}  &  \includegraphics[width=0.49\linewidth]{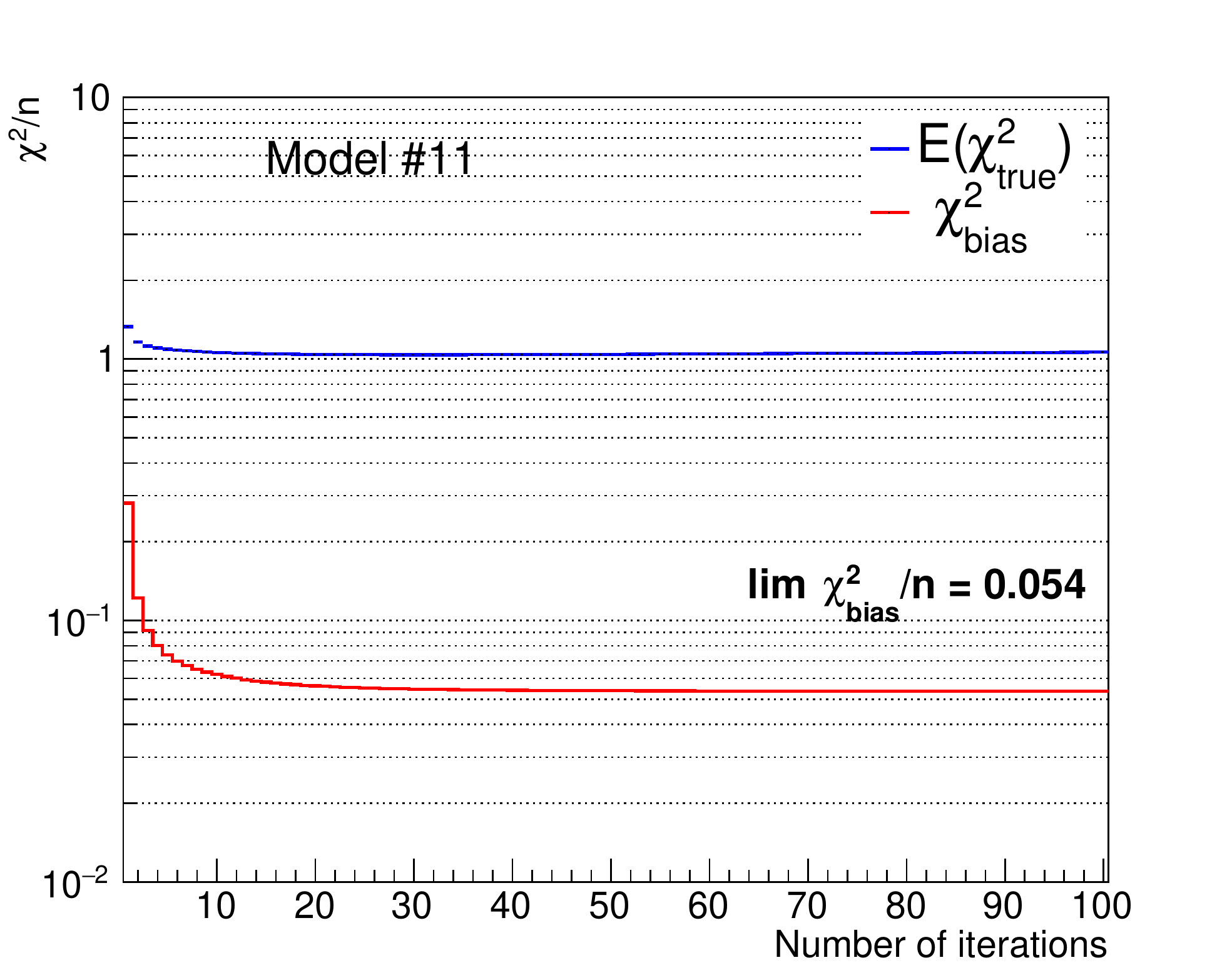}
    \\
    (a) & (b) 
\end{tabular}
\end{center}
	\caption{Evolution of $\chi^2_\mathrm{bias}/n = \mathbb{E}(\chi^2_\mathrm{true}/n) -1 $ when the response matrix is constructed using a smearing with $\sigma_s^\mathrm{MC} = 0.5$ while the true smearing is $\sigma_s = 0.55$. (a) When no systematic uncertainty accounts for this mismodelling, large biases are observed. (b) When a systematic covariance matrix is added in the computation of $\chi^2_\mathrm{bias}$, the relative size of bias with respect to the total uncertainty is significantly reduced.}
	\label{fig:wrongR-syst}
\end{figure}

Let us first mention the most obvious situation where the detector response suffers from systematic uncertainties; it occurs when, for instance, resolution or acceptance are not perfectly known or modelled. This is illustrated in our model when the smearing width $\sigma_s$ used to build the response matrix is different from the real one (cf. figure~\ref{fig:wrongR-syst}), leading to significant biases: $\lim_{k \to \infty} \chi^2_\mathrm{bias}(k)/n \simeq 0.25$. In this context, the addition of a systematic covariance matrix $\mathbf{V}_{k}^\mathrm{syst}$ to the purely statistical $\mathbf{V}_k$: \begin{equation}
    \chi^2_\text{bias}(k) \to \chi^2_\text{bias}(k) = \mathbf{b}_{k}^T \left[\mathbf{V}_{k} + \mathbf{V}_{k}^\mathrm{syst} \right] ^{-1}  \mathbf{b}_{k}
\end{equation} will reduce the relative size of the endpoint bias respectively to the uncertainties and possibly retrieve $\lim_{k \to \infty} \chi^2_\mathrm{bias}(k)/n \ll 1$. In our example (figure~\ref{fig:wrongR-syst} (b)) the bias indeed reduces to $\lim_{k \to \infty} \chi^2_\mathrm{bias}(k)/n \simeq 0.05$.

\begin{figure}[!ht]
	\begin{center}
\begin{tabular}{cc}
   \includegraphics[width=0.49\linewidth]{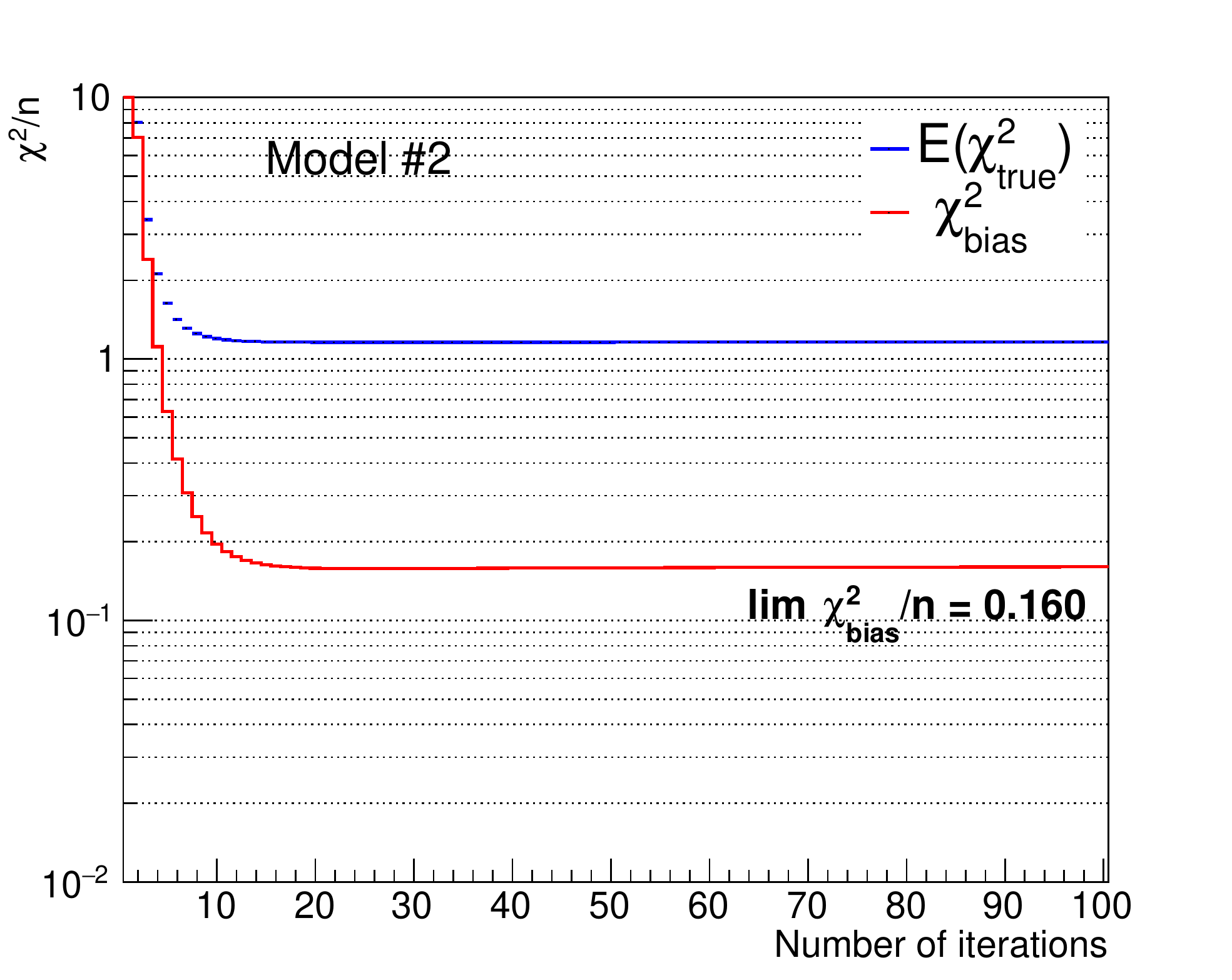}  &  
   \includegraphics[width=0.49\linewidth]{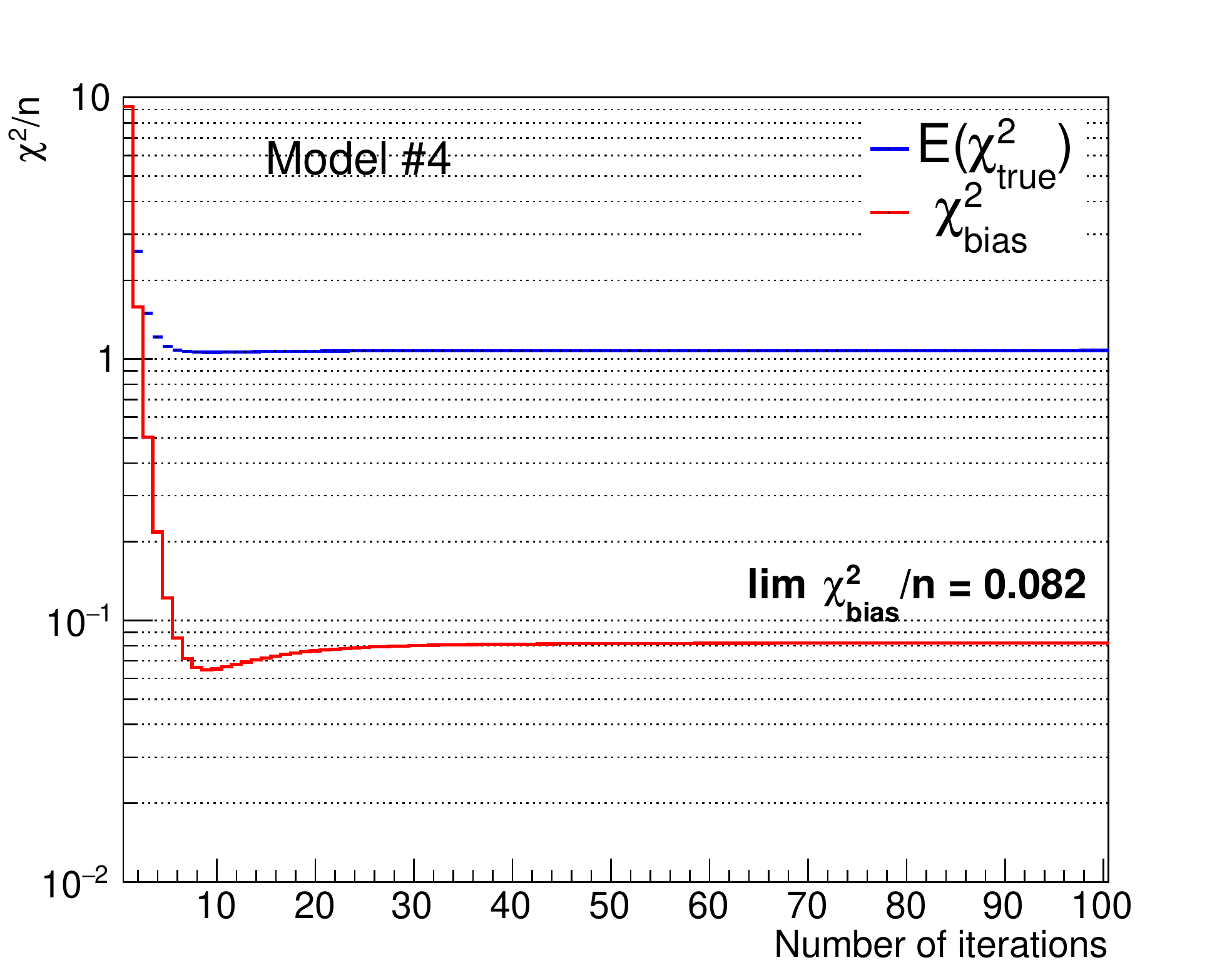} \\
    (a) & (b) 
\end{tabular}
\end{center}
	\caption{Evolution of $\chi^2_\mathrm{bias}/n = \mathbb{E}(\chi^2_\mathrm{true}/n) -1 $ when the true model follows (a) toy model \#2 (shifted peaks) and (b) toy model \#4 (wider second peak). The response matrix built from the nominal model (\#1) has the correct smearing but the distribution $\tilde{\rho}_{X,i}^0(x_t)$ differs from the truth.}
	\label{fig:wrongR-model}
\end{figure}

A second source of inaccuracy comes from the fact that binned distributions are used. Let us note $\rho_{X}(x_t)$ (resp. $\rho_{Y}(y_o)$) the p.d.f. of $X^\mathrm{true}$ (resp. $Y^\mathrm{obs}$). Analytically we have \begin{equation}
    \rho_{Y}(y_o) = \int K (x_t,y_o) \, \rho_{X}(x_t) \, \mathrm{d}x_t
\end{equation} with the kernel $K (x_t,y_o)$ modelling the detector response (a gaussian smearing in our example). The p.d.f. of $Y^\mathrm{obs}$ under the condition that $X^\mathrm{true}$ is in bin $i$ is then
 \begin{equation} \label{eqn:analytic}
    \rho_{Y|X_i^\mathrm{true}}(y_o) = \int_{\mathrm{bin}~i} K (x_t,y_o) \, \tilde{\rho}_{X,i}(x_t) \,  \mathrm{d}x_t
\end{equation} with $\tilde{\rho}_{X,i}(x_t) $ the p.d.f. of $X^\mathrm{true}$ restricted to bin $i$ and normalized so that $\int_{\mathrm{bin}~i} \tilde{\rho}_{X,i} =1$. Finally, the response matrix' coefficient $R_{ij}$ can be expressed as \begin{equation}
    R_{ij} = \int_{\mathrm{bin}~j} \rho_{Y|X_i^\mathrm{true}}(y_o) \, \mathrm{d}y_o = \int_{\mathrm{bin}~j} \int_{\mathrm{bin}~i} K (x_t,y_o) \, \tilde{\rho}_{X,i}(x_t) \,  \mathrm{d}x_t \mathrm{d}y_o.
    \end{equation}
Because true bins $i$ have finite size, the shape of the true distribution $\tilde{\rho}_{X,i}(x_t)$ inside bin $i$ actually matters\footnote{To check the case where true bins have infinitesimal width, let us note the true bin $i$ as $[x_i - \delta, x_i+\delta]$. When $\delta \to 0$, we have $\tilde{\rho}_{X,i}(x_t) \to \delta(x_i - x_t)$ and for all bins $i$: $$\rho_{Y|X_i^\mathrm{true}}(y_o) \to \rho_{Y|X^\mathrm{true}=x_i}(y_o) = K(x_i,y_o).$$ In this case, the conditionnal p.d.f. is determined by the detector response only ($\rho_{Y|X} = K$) and does not depend on the shape of any specific distribution $\rho_X$.}.  To be accurate, one should use the (unknown!) true spectrum to weight events inside a true bin. Instead one only has an educated guess $\tilde{\rho}_{X,i}^0(x_t)$ at best -- not speaking of a flat distribution. As a result the response matrix is inaccurate even when the response kernel $K$ is perfectly known. Figure~\ref{fig:wrongR-model} illustrate this effect: when using alternative distributions $\tilde{\rho}_{X,i}^0(x_t)$ instead of the true one, we obtain $\lim_{k \to \infty} \chi^2_\mathrm{bias}(k)/n \sim 10^{-1}$: the endpoint is biased.

\begin{figure}[!ht]
	\begin{center}
\begin{tabular}{cc}
   \includegraphics[width=0.49\linewidth]{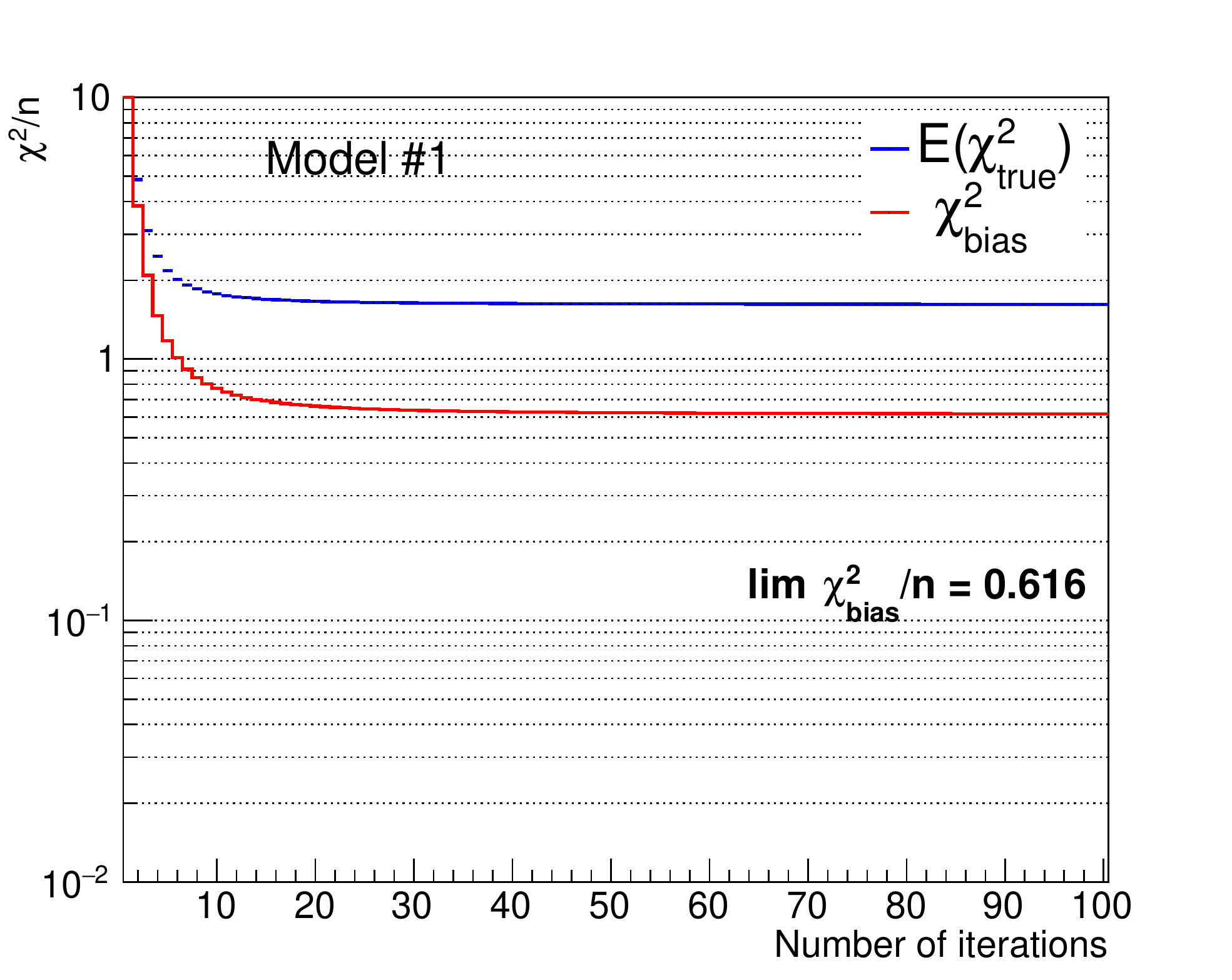}  &  
   \includegraphics[width=0.49\linewidth]{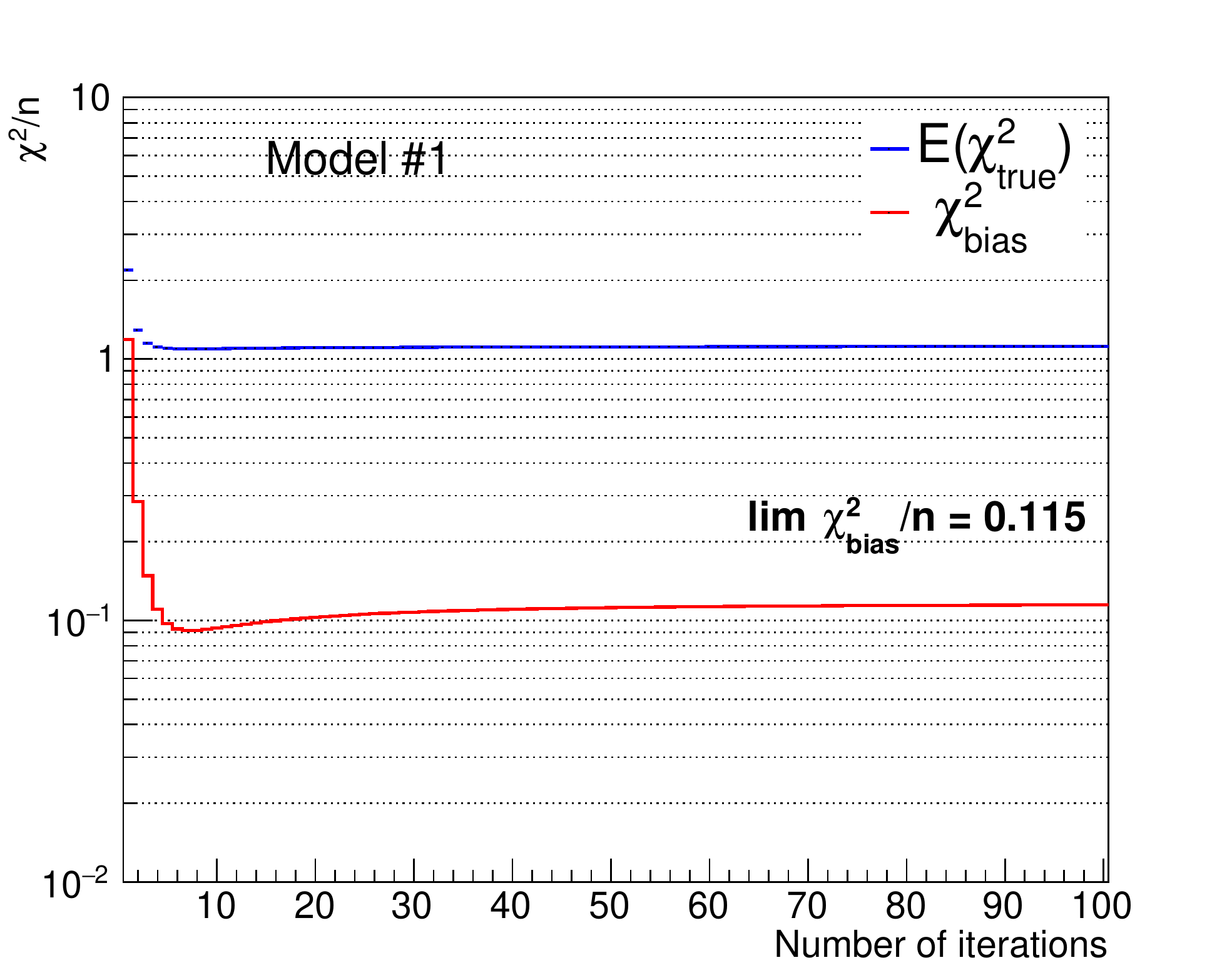} \\
    (a) & (b) 
\end{tabular}
\includegraphics[width=0.49\linewidth]{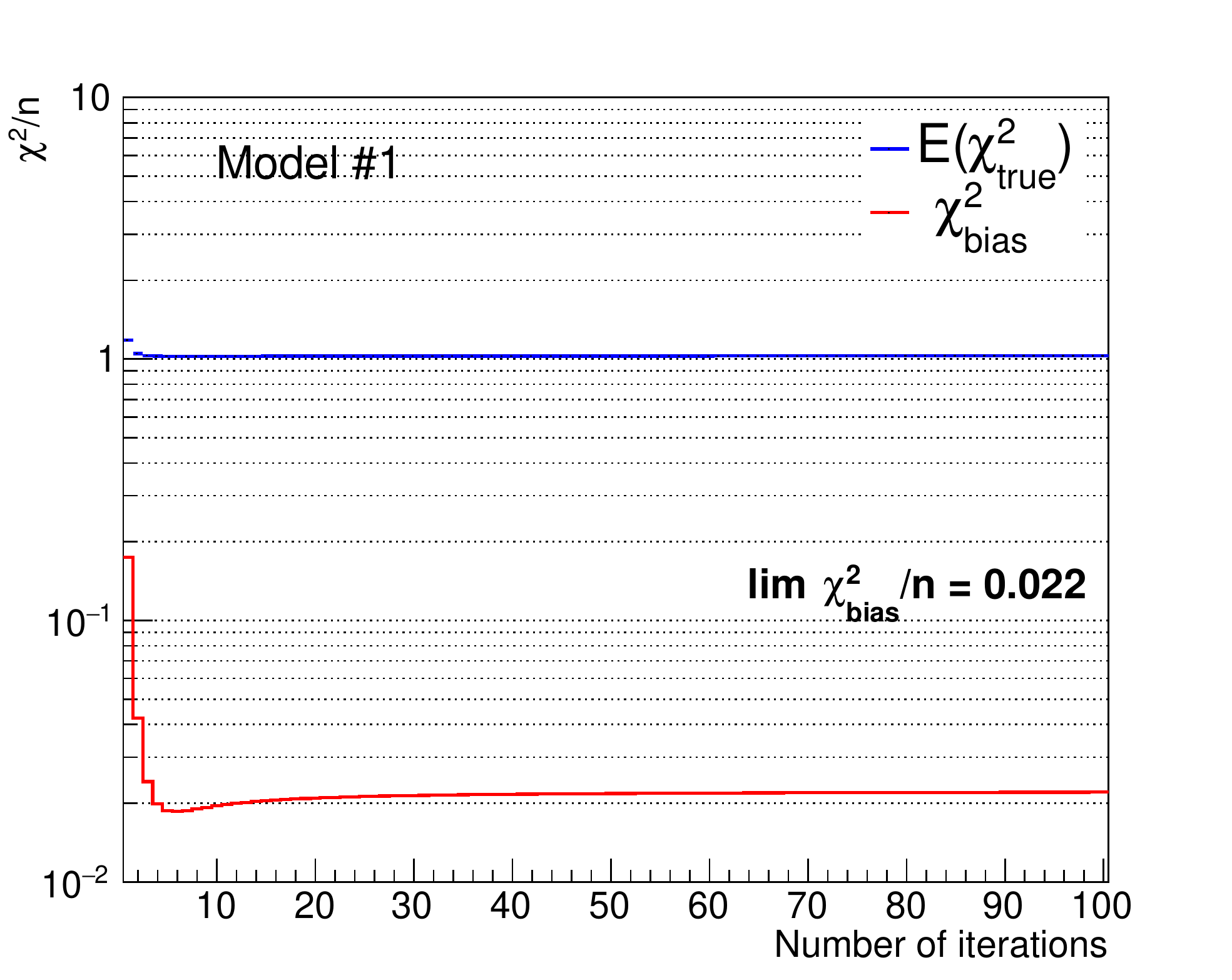}\\
(c)
\end{center}
	\caption{Evolution of $\chi^2_\mathrm{bias}/n = \mathbb{E}(\chi^2_\mathrm{true}/n) -1 $ when the response matrix is sampled from (a) $10^4$ (b) $10^5$ and (c) $10^6$ events. It uses the correct smearing and the true model shape. The true response is generated from an independant sample of $10^6$ events. As expected, lower MC statistics yields higher endpoint bias.}
	\label{fig:wrongR-sampling}
\end{figure}

A third case appears when the response matrix is obtained using MC simulations, which occurs in particular for sophisticated detectors with a complex detector response. The conditional probabilities \begin{equation}
    R_{ij} = \mathbb{P}(Y^\mathrm{obs}_j \; | \; X^\mathrm{true}_i) 
\end{equation} are obtained by sampling particles with $X^\mathrm{true}$ and recording the output quantity $Y^\mathrm{obs}$. Assuming that both the detector response and the true distribution are perfectly known ($K$ and $\tilde{\rho}_X$ in eqn.~(\ref{eqn:analytic})), limited sample size will still blur the response matrix. Even when large samples are accessible, the response matrix is never \textit{exact} strictly speaking. Figure~\ref{fig:wrongR-sampling} provides an example with our nominal model, sampling the response matrix with $10^4$, $10^5$ or events $10^6$.  We obtain for $\lim_{k \to \infty} \chi^2_\mathrm{bias}(k)/n $ values of about 0.6, 0.1, and 0.02 respectively: depending on sample size, the bias may, or may not, be negligible. As expected, the larger the sample, the smaller the bias; but the bias is there in all cases.

In summary, several sources of response matrix inaccuracy exist: systematic uncertainties on the detector response, finite bin size in true space, sampling of the response matrix. Hence, endpoint biases are present and the unregularized spectrum $\hat{\mathbf{N}}_\infty$ is not an unbiased estimator. However, it may occur that the endpoint bias $\mathbf{b}_\infty$ is actually negligible when compared to uncertainties ($\lim_{k \to \infty} \chi^2_\mathrm{bias}(k)/n \ll 1 $). We therefore suggest that analyzers investigate this point in their own context and quantify this intrinsic bias. We believe that systematic uncertainties are well treated in most analyses; however, sampling and model shape issues may not be considered in general. One possible way to study these effects is to build alternative response matrices from, e.g., an independent MC sample or a different event distribution $\tilde{\rho}_{X,i}(x_t)$ and to quantify the induced discrepancy on unfolded spectra. If not negligible, systematic uncertainties may need to be assigned to the response matrix' construction.\\

\begin{figure}[!ht]
	\begin{center}
\begin{tabular}{cc}
   \includegraphics[width=0.49\linewidth]{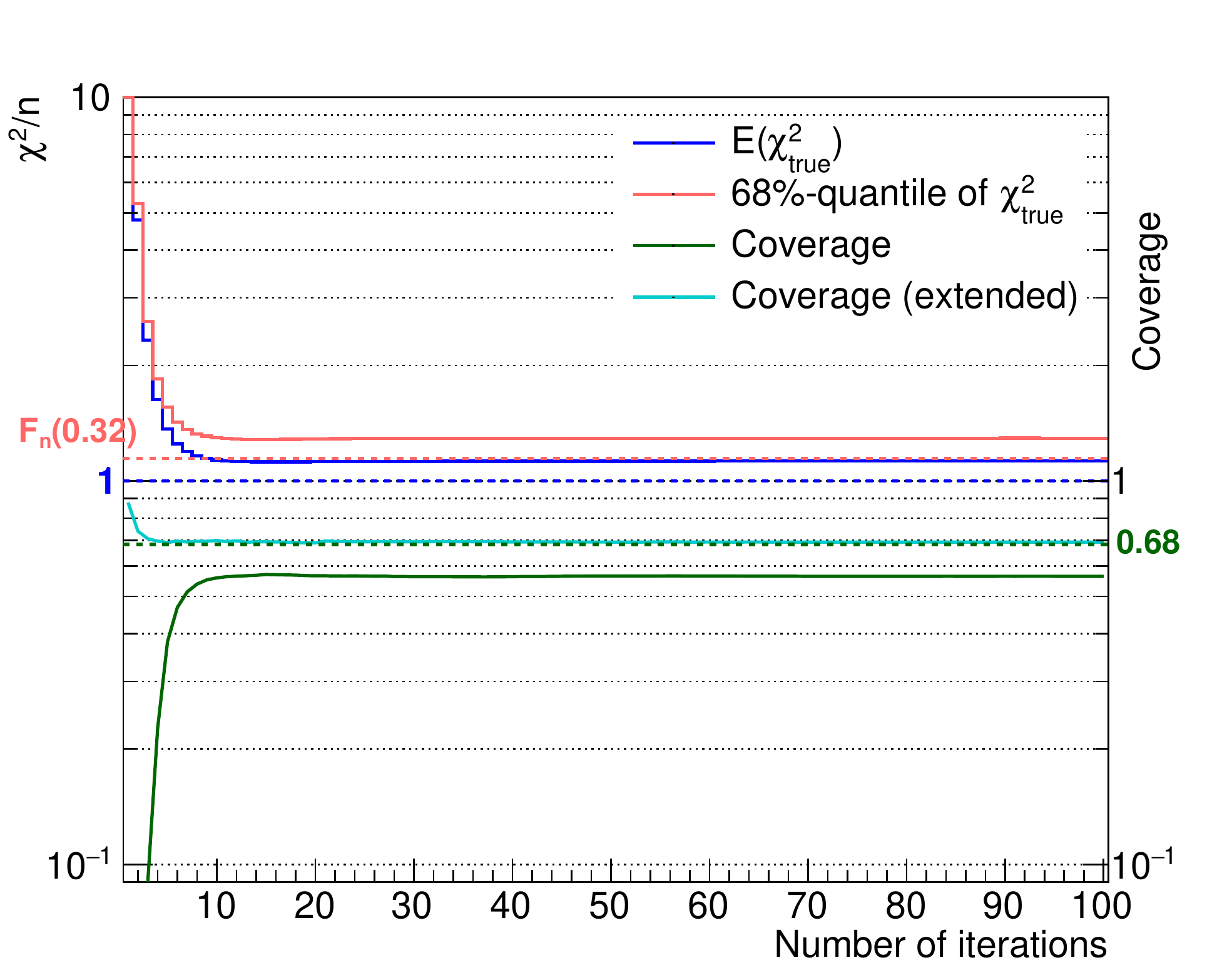}  &  
   \includegraphics[width=0.49\linewidth]{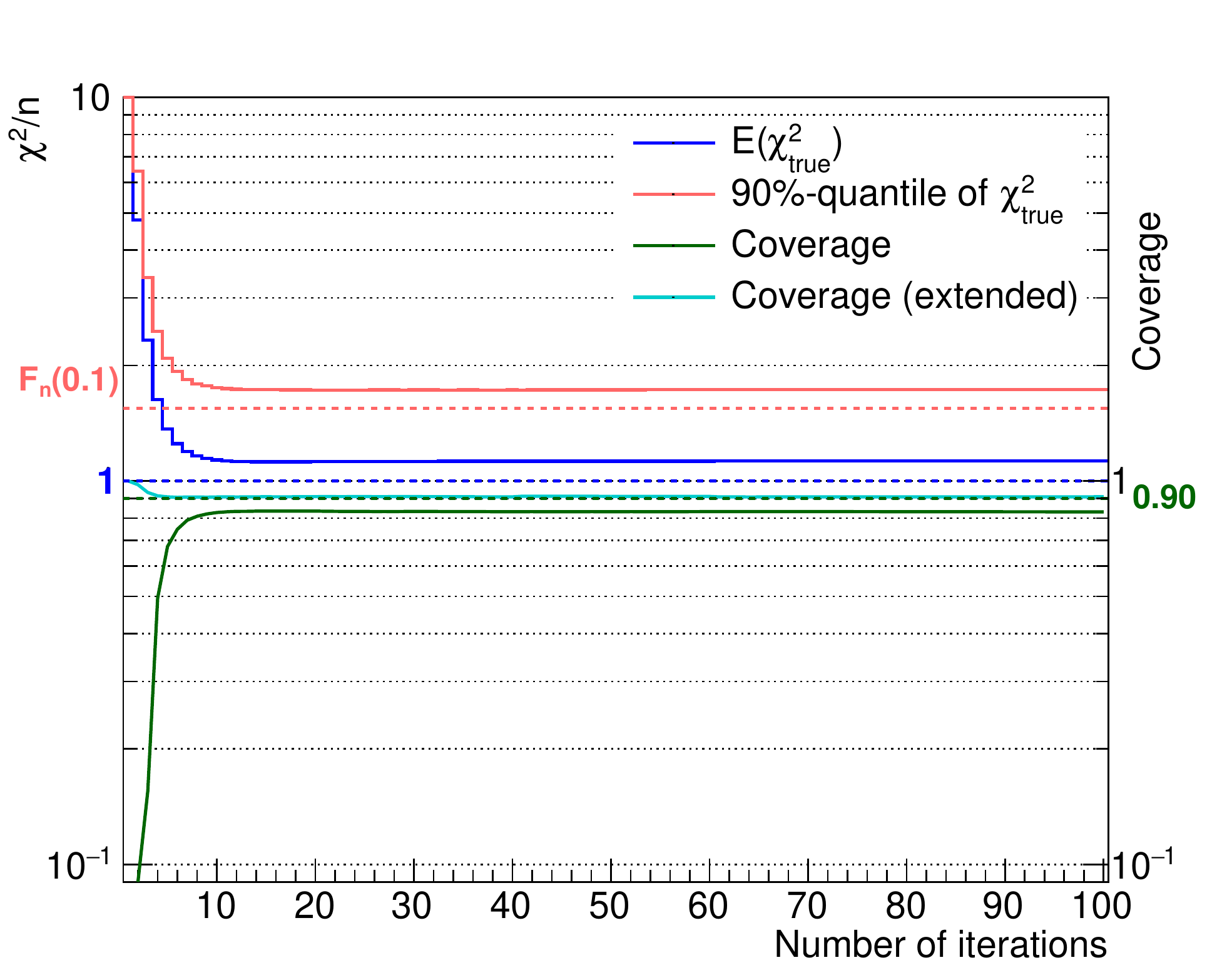} \\
    (a) & (b) 
\end{tabular}
\end{center}
	\caption{Evolution of coverage of confidence regions for (a) 68\% CL and (b) 90\% CL, produced using model \#3. An inexact response matrix is used, leading to $\lim_{k \to \infty} \mathbb{E}[\chi^2_\mathrm{true}(k)]/n = 1.17$ (biased endpoint). The 68\%-quantile of $\chi^2_\mathrm{true}$ is the value of $\chi^2_\text{lim}$ that provides a 68\% CL confidence region with proper coverage; it corresponds to $F_n (1-\alpha)$ (red dotted line) in the no-bias limit, which is never reached in this case. The extended confidence region defined by $\chi^2_\text{lim}(\alpha; \chi^2_\mathrm{bias})$ achieves correct coverage even in presence of large biases.}
	\label{fig:coverage-wrongR}
\end{figure}

As for the data-driven criterion presented in this article, the control of the bias provided by $\chi^2_\mathrm{bias}/n \leqslant \varepsilon^2$ does not depend on the nature of the bias, either a truncation bias or an endpoint bias. As a result, it remains applicable for inaccurate response matrices. An example of coverage evolution is given in figure~\ref{fig:coverage-wrongR}. It is the equivalent of figure~\ref{fig:coverage}, but the response matrix is not accurate anymore: it is sampled from $10^6$ events, using the nominal model (\#1) instead of the true one (here model \#3). The confidence region defined with the extended boundary $\chi^2_\mathrm{lim}(\alpha,\chi^2_\mathrm{bias})$ from eqn.~(\ref{eqn:confidence-region-size-ext}) allows to recover appropriate coverage in this case as well.

\begin{figure}[!ht]
	\begin{center}
\begin{tabular}{cc}
   \includegraphics[width=0.49\linewidth]{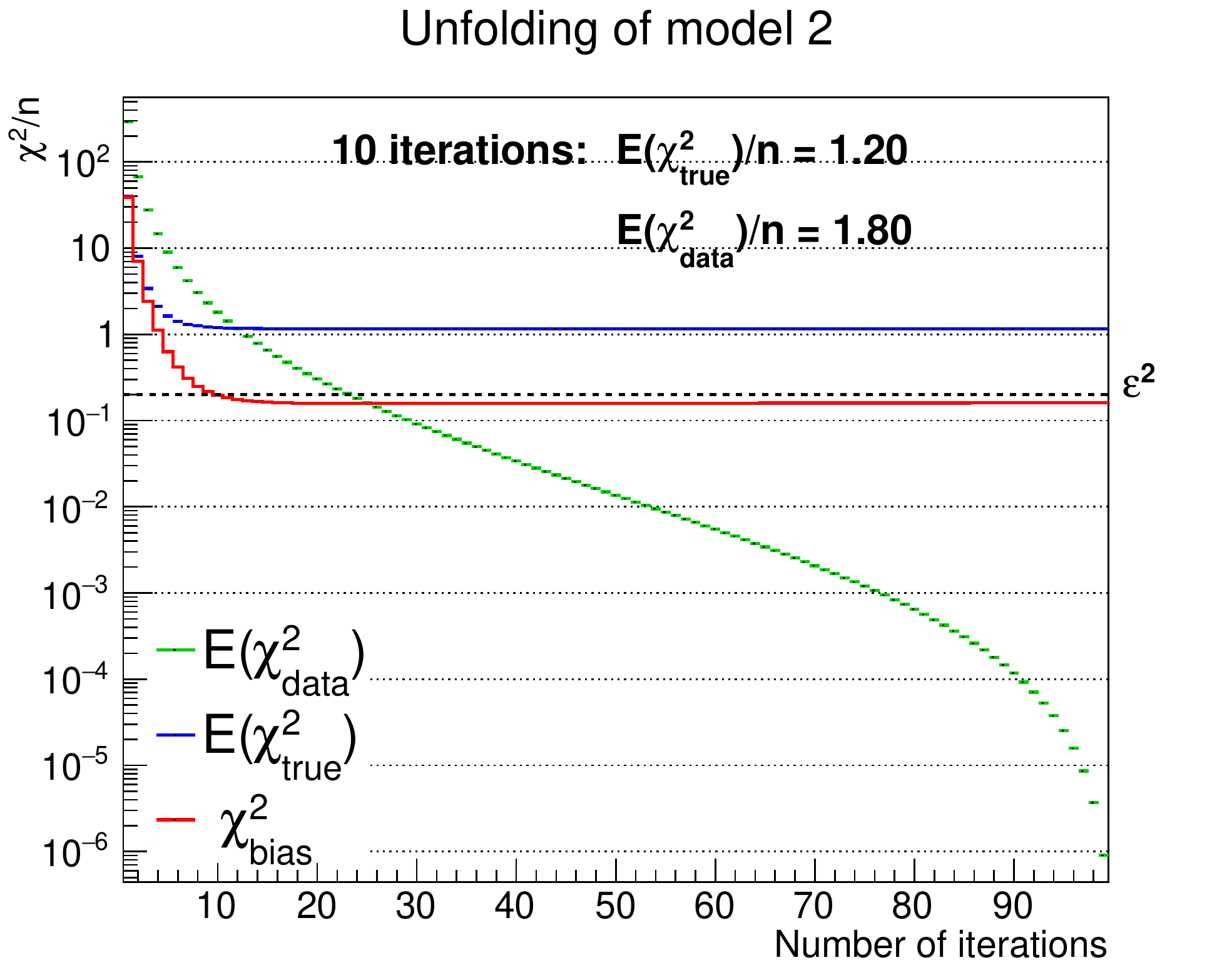}  &
   \includegraphics[width=0.49\linewidth]{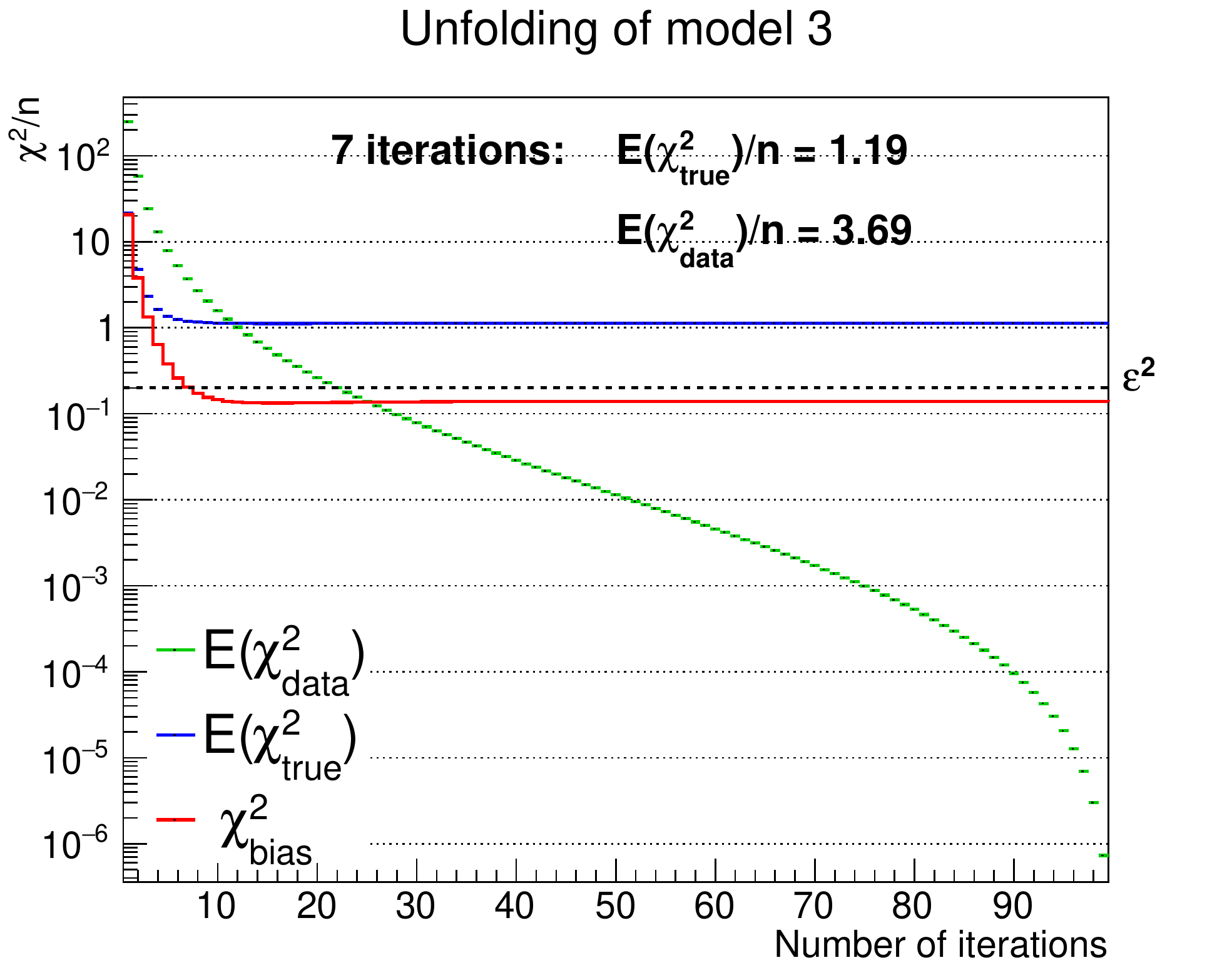} \\
   \includegraphics[width=0.49\linewidth]{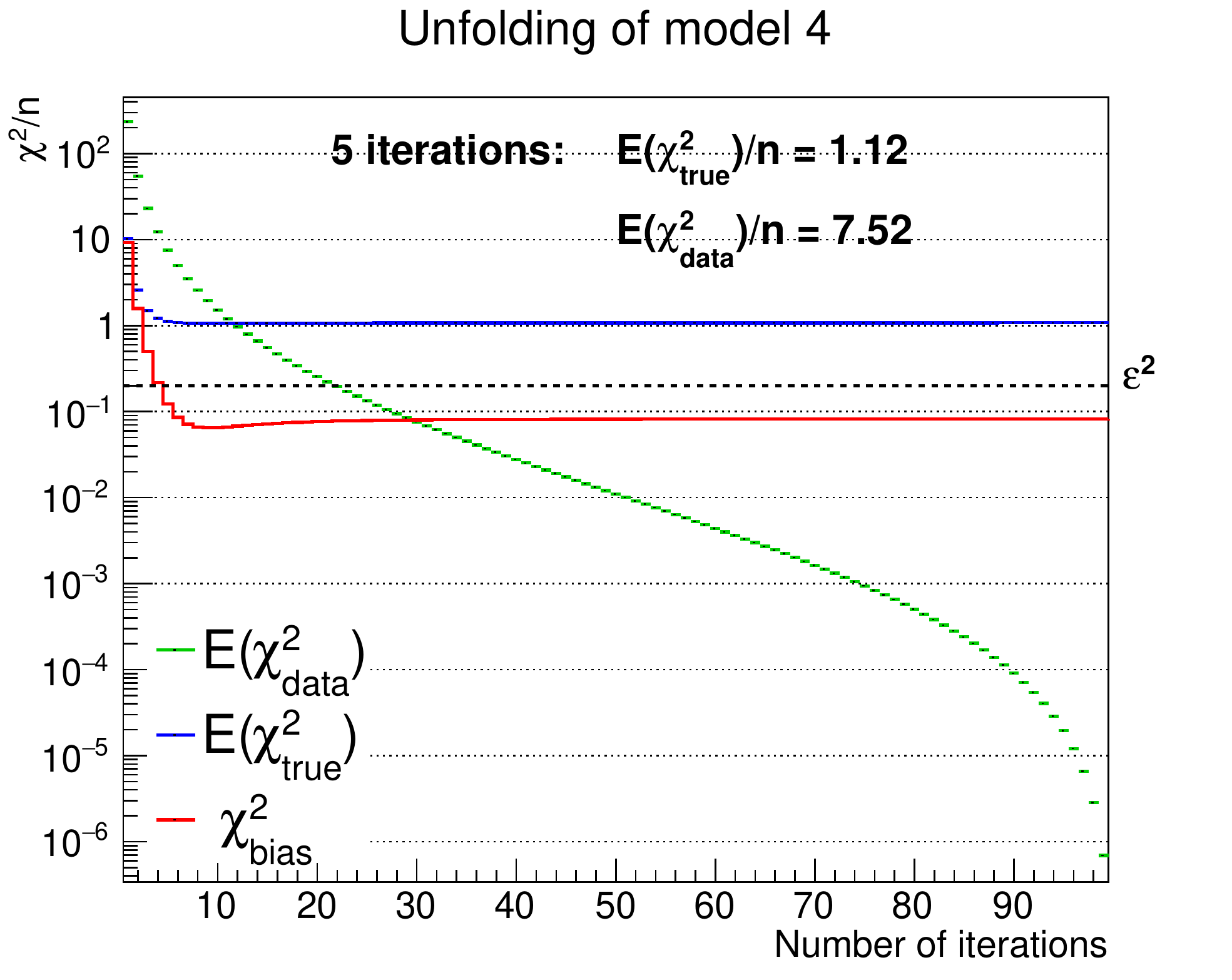}  &
   \includegraphics[width=0.49\linewidth]{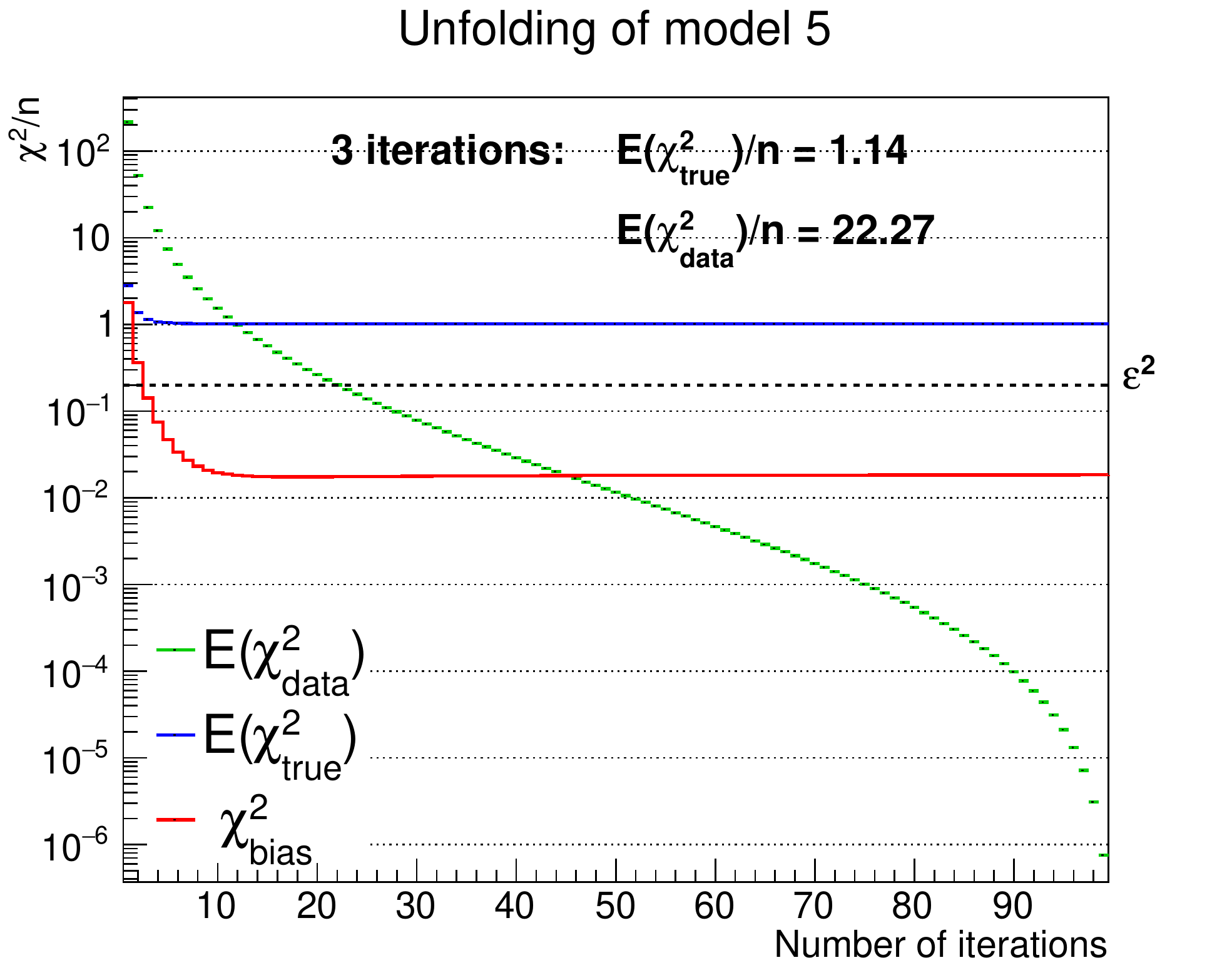}  \\
    \includegraphics[width=0.49\linewidth]{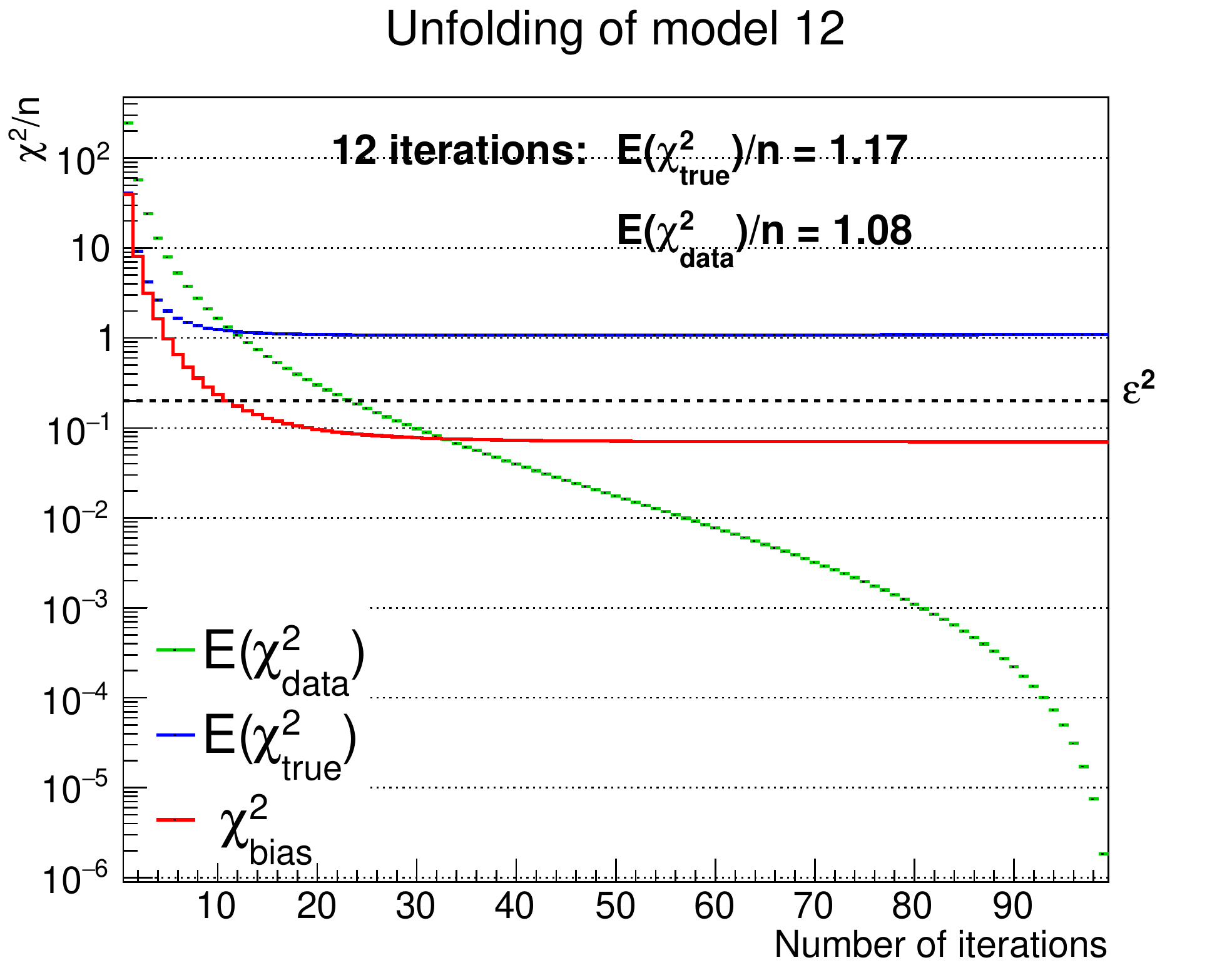}  &
   \includegraphics[width=0.49\linewidth]{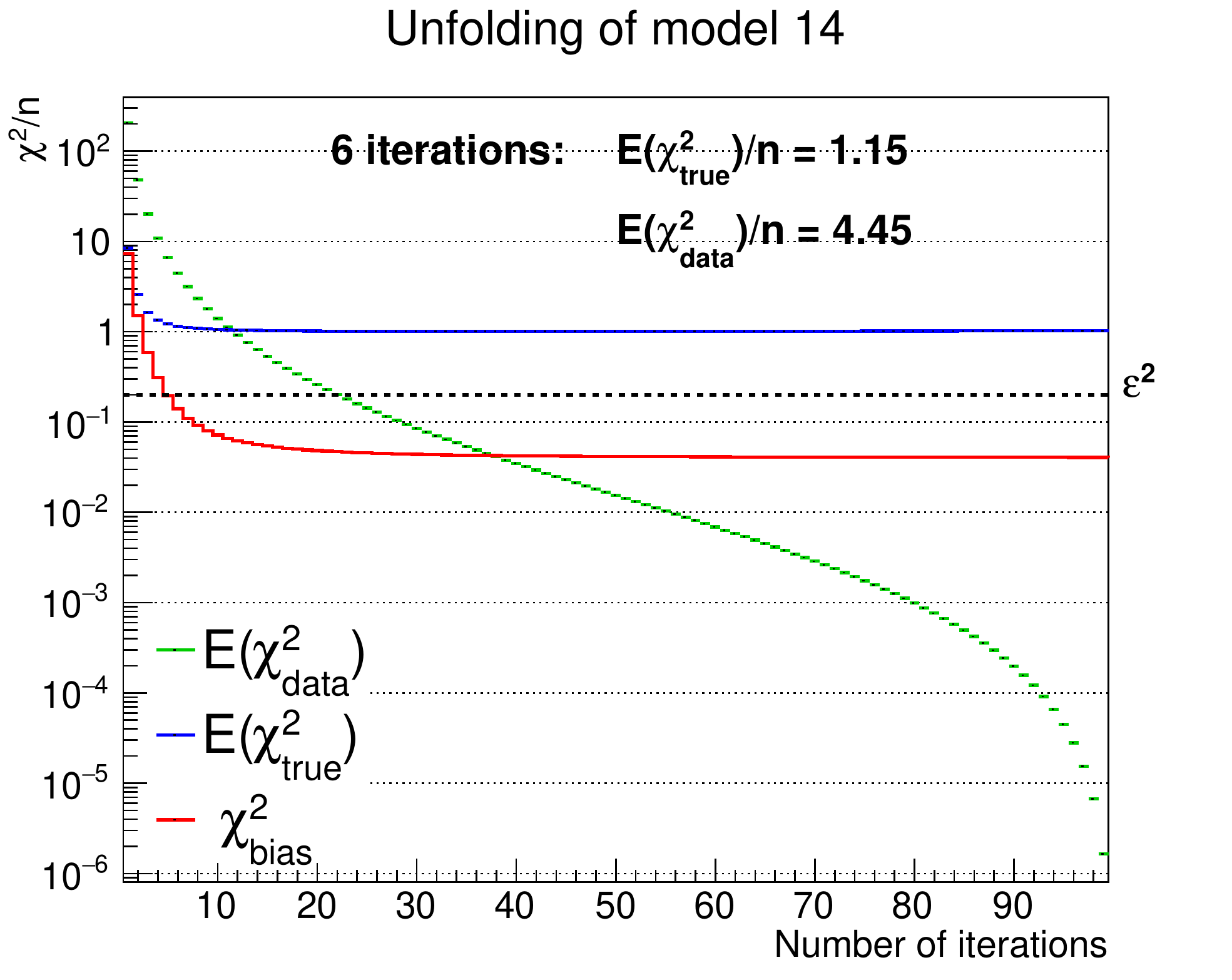}  \\
\end{tabular}
\end{center}
	\caption{Evolution of $\mathbb{E}(\chi^2_\mathrm{data}/n)$ and $\chi^2_\mathrm{bias}/n = \mathbb{E}( \chi^2_\mathrm{true}/n)-1$ for toy model \#2-5. For each model, the response is sampled using model \#1, leading to small but non-zero endpoint bias. For models \#12 and \#14, having a smearing $\sigma_s$ different from the nominal, a systematic covariance is computed and included in the $\chi^2$ definitions.
	The first iteration such that $\mathbb{E}(\chi^2_\mathrm{bias}/n) \leqslant \varepsilon^2$ is indicated, with $\varepsilon^2 = 0.2$. The corresponding values of $\mathbb{E}(\chi^2_\mathrm{data}/n)$ are indicated as well; $\eta^2$ is set as the lowest $\mathbb{E}(\chi^2_\mathrm{data}/n)$ among all models.}
	\label{fig:chi2-data}
\end{figure}

Although the endpoint spectrum is not an unbiased MLE anymore, it remains prior-independent. Therefore, the benefits of the criterion presented in section~\ref{scn:DD-criterion}, based on $\chi^2_\mathrm{data}$, are still relevant. The value of $\eta^2$ satisfying the convergence condition~(\ref{eqn:criterion2}) can be obtained using pseudo-data studies with alternative true models, as in section \ref{scn:DD-illustration}. The results are displayed on figure~\ref{fig:chi2-data}; models requiring a systematic uncertainty related to the smearing are also considered. The worst-case scenario is taken to set $\eta^2=1.08$.

\section{Discussion and summary}\label{scn:ccl}
The D'Agostini (Richardson-Lucy) iterative unfolding has developed to become one of the most frequently used unfolding technique. To the best of our knowledge, only a few elaborated methods to set the number of iterations have been published in the specific context of this algorithm. Most notably, we mention here the method by G. Zech \cite{Zech_2013}, where acceptable numbers of iterations $k$ are such that $\hat{\mathbf{N}}_{k}$ fits the data almost as well as the unregularized best-fit $\hat{\mathbf{N}}_{\infty}$. The observable of interest is $\Delta \chi^2(k) = \chi^2(\hat{\mathbf{N}}_{k}) - \chi^2(\hat{\mathbf{N}}_\infty)$ (or the corresponding $p$-value), on which a threshold is set. As in our method, the endpoint spectrum is taken as reference; however, $\chi^2_\mathrm{data}$ compares directly $\hat{\mathbf{N}}_{k}$ to $\hat{\mathbf{N}}_{\infty}$ instead of their respective goodness-of-fit with data. We believe these are complementary approaches.

Concerning uncertainty quantification, ref.~\cite{Zech_2013} recommends to provide the unregularized covariance matrix $\mathbf{V}_\infty$, which is meant to ensure proper coverage for the corresponding confidence intervals. However, as showed in section~\ref{scn:wrong-response}, the true response matrix is in most cases not perfectly known. Thus, the endpoint spectrum $\hat{\mathbf{N}}_\infty$ remains a biased estimator, and the covariance $\mathbf{V}_\infty$ actually undercovers. 

Another interesting idea for uncertainty quantification has been proposed by M. Kuusela and V. Panaretos \cite{Kuusela:2015xqa}, although not in the specific context of D'Agostini unfolding. It is called \textit{bias-corrected uncertainty quantification}. The spectrum estimate is chosen with a generally strong regularization, but the corresponding covariance is iteratively corrected for the regularization bias until desired coverage is retrieved. Within the iterative unfolding framework discussed here, this would translate into picking a $k_0$ for the spectrum estimator $\hat{\mathbf{N}}_{k_0}$, and a larger $k_1$ to have a less regularized covariance $\mathbf{V}_{k_1}$. However, in presence of endpoint biases, appropriate coverage may be beyond the reach of the unfolding, at any iteration. 

Controlling the amount of bias is a key point for analyses. The convergence criterion presented in this article allows to control the level of bias introduced (by setting $\varepsilon^2$) and suggests a way to extend confidence regions to retrieve the expected coverage (by using $\chi^2_\mathrm{lim}(\alpha;n\varepsilon^2)$). As the statistical variance inflates with the number of iterations, a well-controlled covariance $\mathbf{V}_k$ also provides smaller, yet meaningful, error bars than the unregularized $\mathbf{V}_\infty$.

In addition, with the convergence criterion presented here, the stopping iteration is not determined \textit{a priori}. This is particularly relevant when the true model is suspected to not be well reproduced by MC generators. In such cases, MC-based studies may fail to evaluate or control the level of bias. With this new method, we aim at being as much data-driven as possible, while providing valid uncertainties, which are crucial for model comparisons using unfolded data.

\section*{Appendix}

\subsection*{Proof of relation~(\ref{eqn:E(chi2true)})}Let us denote is $\Delta \mathbf{N}_{k} \equiv \hat{\mathbf{N}}_{k}  - \overline{\mathbf{N}}$, and $\mathbf{b}_k \equiv \langle \Delta \mathbf{N}_{k}  \rangle$; $\chi^2_\mathrm{true}$ can be rewritten as\begin{equation}
        \chi^2_\mathrm{true}(k) = \Delta \mathbf{N}_{k}^T \cdot \mathbf{V}_{k}^{-1} \cdot \Delta \mathbf{N}_{k}.
    \end{equation} As a scalar number, $\chi^2_\mathrm{true}(k)$ equals its trace and \begin{equation}
        \chi^2_\mathrm{true}(k) = \text{Tr} \left[\Delta \mathbf{N}_{k}^T \mathbf{V}_{k}^{-1}  \Delta \mathbf{N}_{k} \right] = \text{Tr} \left[\mathbf{V}_{k}^{-1} \Delta \mathbf{N}_{k} \Delta \mathbf{N}_{k}^T  \right].
    \end{equation}
       The expectation value of $\Delta \mathbf{N}_{k} \Delta \mathbf{N}_{k}^T$ is by definition the covariance with respect to the true spectrum $\mathbf{V}_{T,k}$: \begin{equation} \mathbf{V}_{T,k} \equiv \mathbb{E}\big[( \mathbf{N}_{k} - \overline{\mathbf{N}}) ( \mathbf{N}_{k} - \overline{\mathbf{N}})^T \big] = \mathbf{V}_{k} +  \mathbf{b}_{k} \mathbf{b}_{k}^T. \end{equation} In this bias-variance decomposition, $\mathbf{V}_{k}$ corresponds to the variance of $\hat{\mathbf{N}}_k$ around its expectation value and $ \mathbf{b}_{k} \mathbf{b}_{k}^T$ accounts for the average bias. Using the linearity of the trace operator and of the expectation value, we get \begin{equation}
            \mathbb{E}\big[\chi^2_\mathrm{true}(k)\big] = \text{Tr} \left[\mathbf{V}_{k}^{-1}\,  \mathbb{E}\big(\Delta \mathbf{N}_{k} \Delta \mathbf{N}_{k}^T \big)  \right] = \text{Tr} \left[\mathbf{V}_{k}^{-1} \mathbf{V}_{T,k} \right] 
       \end{equation} yielding \begin{equation}         \begin{split}
            \mathbb{E}\big[\chi^2_\mathrm{true}(k)\big] & = \text{Tr} \left[\mathbb{I}_n + \mathbf{V}_{k}^{-1}  \mathbf{b}_{k} \mathbf{b}_{k}^T \right] \\
        & = n + \chi^2_\text{bias}(k).
    \end{split} 
    \end{equation}

\bibliography{mybibfile}

\begin{thebibliography}{10}
\expandafter\ifx\csname url\endcsname\relax
  \def\url#1{\texttt{#1}}\fi
\expandafter\ifx\csname urlprefix\endcsname\relax\def\urlprefix{URL }\fi
\expandafter\ifx\csname href\endcsname\relax
  \def\href#1#2{#2} \def\path#1{#1}\fi

\bibitem{1972Richardson}
W.~H. {Richardson}, {Bayesian-Based Iterative Method of Image Restoration},
  Journal of the Optical Society of America 62 (1972) 55.

\bibitem{Lucy:1974yx}
L.~Lucy, {An iterative technique for the rectification of observed
  distributions}, Astron. J. 79 (1974) 745--754.
\newblock \href {https://doi.org/10.1086/111605} {\path{doi:10.1086/111605}}.

\bibitem{Multhei:1986ps}
H.~Multhei, B.~Schorr, {On an Iterative Method for the Unfolding of Spectra},
  Nucl. Instrum. Meth. A 257 (1987) 371.
\newblock \href {https://doi.org/10.1016/0168-9002(87)90759-5}
  {\path{doi:10.1016/0168-9002(87)90759-5}}.

\bibitem{H_cker_1996}
A.~Höcker, V.~Kartvelishvili, {SVD approach to data unfolding}, Nucl. Instrum.
  Meth. A 372 (1996) 469–481.
\newblock \href {https://doi.org/10.1016/0168-9002(95)01478-0}
  {\path{doi:10.1016/0168-9002(95)01478-0}}.

\bibitem{DAgostini1995}
G.~D'Agostini, {A Multidimensional unfolding method based on Bayes' theorem},
  Nucl. Instrum. Meth. A 362 (1995) 487--498.
\newblock \href {https://doi.org/10.1016/0168-9002(95)00274-X}
  {\path{doi:10.1016/0168-9002(95)00274-X}}.

\bibitem{blobel2002unfolding}
V.~Blobel, {An Unfolding Method for High Energy Physics Experiments} (2002).
\newblock \href {http://arxiv.org/abs/hep-ex/0208022}
  {\path{arXiv:hep-ex/0208022}}.

\bibitem{Lcurve}
P.~C. Hansen, {Analysis of Discrete Ill-Posed Problems by Means of the
  L-Curve}, SIAM Review 34 (1992) 561--580.

\bibitem{GCV}
G.~Wahba, G.~H. Golub, M.~Heath, {Generalized Cross-Validation as a Method for
  Choosing a Good Ridge Parameter}, Technometrics, Vol. 21, no 2 (1979).

\bibitem{Tikhonov:1963}
A.~N. Tikhonov, Solution of incorrectly formulated problems and the
  regularization method, Soviet Math. Dokl. 4 (1963) 1035--1038.

\bibitem{TSVD}
P.~C. Hansen, {The truncated SVD as a method for regularization}, BIT 27 (1987)
  534--553.
\newblock \href {https://doi.org/10.1007/BF01937276}
  {\path{doi:10.1007/BF01937276}}.

\bibitem{WienerSVD}
W.~Tang, X.~Li, X.~Qian, H.~Wei, C.~Zhang, {Data Unfolding with Wiener-{SVD}
  Method}, JINST 12 (2017) P10002--P10002.
\newblock \href {https://doi.org/10.1088/1748-0221/12/10/p10002}
  {\path{doi:10.1088/1748-0221/12/10/p10002}}.

\bibitem{Zech_2013}
G.~Zech, {Iterative unfolding with the Richardson–Lucy algorithm}, Nucl.
  Instrum. Meth. A 716 (2013) 1–9.
\newblock \href {https://doi.org/10.1016/j.nima.2013.03.026}
  {\path{doi:10.1016/j.nima.2013.03.026}}.

\bibitem{zech2016analysis}
G.~Zech, Analysis of distorted measurements -- parameter estimation and
  unfolding (2016).
\newblock \href {http://arxiv.org/abs/1607.06910} {\path{arXiv:1607.06910}}.

\bibitem{DAgostini2010}
G.~D'Agostini, {Improved iterative Bayesian unfolding}, in: {Alliance Workshop
  on Unfolding and Data Correction}, 2010.
\newblock \href {http://arxiv.org/abs/1010.0632} {\path{arXiv:1010.0632}}.

\bibitem{KuuselaMSc}
M.~J. Kuusela, {Statistical Issues in Unfolding Methods for High Energy
  Physics. Master's thesis, Aalto University} (2012).

\bibitem{licci2018}
M.~Licciardi, {Etude de la production d’un pion dans l’interaction de
  neutrinos muoniques avec le nouveau d\'etecteur WAGASCI au Japon. PhD thesis,
  Université Paris-Saclay} (2018).

\bibitem{NEUT}
Y.~Hayato, {A neutrino interaction simulation program library NEUT}, Acta Phys.
  Polon. B 40 (2009) 2477--2489.

\bibitem{GENIE}
C.~Andreopoulos, {The GENIE neutrino Monte Carlo generator}, Acta Phys. Polon.
  B 40 (2009) 2461--2475.

\bibitem{KuuselaPhD}
M.~J. Kuusela, {Uncertainty quantification in unfolding elementary particle
  spectra at the Large Hadron Collider. PhD thesis, EPFL} (2016).
\newblock \href {https://doi.org/10.5075/epfl-thesis-7118}
  {\path{doi:10.5075/epfl-thesis-7118}}.

\bibitem{adye2011}
T.~Adye, {Unfolding algorithms and tests using RooUnfold} (2011).
\newblock \href {http://arxiv.org/abs/1105.1160} {\path{arXiv:1105.1160}}.

\bibitem{PDGstatistics}
{P. A. Zyla et al. (Particle Data Group)}, {Statistics}, {Prog. Theor. Exp.
  Phys. 2020, 083C01 (2020)}.

\bibitem{Kuusela:2015xqa}
M.~Kuusela, V.~M. Panaretos, {Statistical unfolding of elementary particle
  spectra: Empirical Bayes estimation and bias-corrected uncertainty
  quantification}, Ann. Appl. Stat. 9 (2015) 1671--1705.
\newblock \href {https://doi.org/10.1214/15-AOAS857}
  {\path{doi:10.1214/15-AOAS857}}.

\end{thebibliography}

\end{document}